\documentclass[a4paper,11pt]{article}

\usepackage{jheppub1} 

\usepackage[T1]{fontenc} 
\usepackage{empheq,mathtools}
\usepackage{cancel,slashed}

\def\arXiv#1{\href{http://arxiv.org/abs/#1}{arXiv:#1}}
\def\arXiv#1#2{\href{http://arxiv.org/abs/#1}{arXiv:#1}}
\def\arXivid#1#2{\href{http://arxiv.org/abs/#1/#2}{#1/#2}}
\def\doi#1#2{\href{http://doi.org/#1}{#2}}

\title{\boldmath Thermodynamics of hairy accelerating black holes in gauged supergravity and beyond}

\author{Yi Wang and Jie Ren}

\affiliation{School of Physics, Sun Yat-sen University, Guangzhou, 510275, China}

\emailAdd{wangy973@mail2.sysu.edu.cn}
\emailAdd{renjie7@mail.sysu.edu.cn}

\abstract{We study the thermodynamics of accelerating asymptotically AdS black holes with scalar hair in four-dimensional Einstein-Maxwell-dilaton theories, which can be embedded in gauged supergravities for specific dilaton coupling constants. These accelerating black holes are described by charged dilaton C metrics, which have unique features but are almost unexplored before. Thermodynamics of slowly accelerating black holes is generalized to include scalar hair. We find that properly accounting for mixed boundary conditions for the scalar field leads to standard consistent first law and other thermodynamic relations. We compute the dual stress-energy tensor and the mass through holographic renormalization, which is associated with the boundary conditions of the scalar field. We also find that the conformal mass is the same as the holographic mass. Along the same line, we also obtain the thermodynamics of $U(1)^2$-charged accelerating black holes.}

\begin{document} 
\maketitle
\flushbottom

\section{Introduction}

The C metric is a class of exact solutions of Einstein field equations, and the original one was found by Hermann Weyl in 1917 \cite{Weyl:1917ur}. This name came from the C slot of the classification by Ehlers and Kundt in 1963 \cite{EhlersKundt}. A breakthrough work by Kinnersley and Walker in 1970 \cite{Kinnersley:1970zw} reveals that this solution describes a pair of accelerating black holes apart from each other. This solution was generalized to include rotation and the cosmological constant by Plebański-Demiański in 1976 \cite{Plebanski:1976gy}. For more interpretations of this solution, see \cite{Dias:2002mi, Griffiths:2006tk} for example. Based on the AdS/CFT correspondence, the C metric has fruitful applications, for instance, the construction of exact black holes on the branes \cite{Emparan:1999wa, Emparan:1999fd}, the plasma ball \cite{Emparan:2009dj}, the black funnels and droplets \cite{Hubeny:2009kz, Caldarelli:2011wa}, the quantum BTZ black hole \cite{Emparan:2020znc}, and the spinning spindles from accelerating black holes \cite{Ferrero:2020twa}.

Black holes are perhaps the most intriguing objects in our Universe, connecting the bridge between classical and quantum gravity. Especially this link is highlighted by the early discoveries that black hole as a thermal object obeys the laws of thermodynamics \cite{Bekenstein:1973ur, Bekenstein:1974ax, Hawking:1974sw}, and the black hole chemistry \cite{Kubiznak:2016qmn} gives more diversity to these thermal objects. Although the study of the C metric has a long history, the thermodynamics of the accelerating black holes has been formulated for small acceleration in AdS spacetime in recent years \cite{Astorino:2016ybm, Astorino:2016xiy, Appels:2016uha, Appels:2017xoe}. Later, the authors of \cite{Anabalon:2018ydc, Anabalon:2018qfv} formulated extended thermodynamics for slowly accelerating black holes in AdS spacetime, in which it is even allowed to vary the tension along each axis independently. A key ingredient of their works is a rescaling parameter $\alpha$ for the time coordinate as pointed out before in \cite{Gibbons:2004ai} for Kerr-anti-de Sitter (AdS) black holes. Last year, the authors of \cite{Cassani:2021dwa} generalized previous results to include magnetic charge and allow the variation of Newton's constant in the first law of thermodynamics. From the holographic perspective, varying Newton's constant together with the AdS radius can keep the corresponding conformal field theory (CFT) unchanging \cite{Visser:2021eqk, Cong:2021fnf}.

More generally, asymptotically AdS black holes with scalar hair in gauged supergravity are valuable in the AdS/CFT correspondence. The charged dilaton C metrics found in \cite{Lu:2014ida, Ren:2019lgw} are significant generalizations of the original C metric. They describe asymptotically AdS accelerating black holes with scalar hair in Einstein-Maxwell-dilaton systems. The scalar potential of these systems can be expressed by a superpotential, and these theories can be embedded in gauged supergravities for specific dilaton coupling constants. Alternatively, these special case solutions can be uplifted on $S^7$ to $D=11$ supergravity (see Appendix A for more details), and an uplifting AdS$_4\times S^7$ solution dual to a $d=3$ supersymmetric conformal field theories on the boundary was studied in \cite{Ferrero:2020twa}. Even without acceleration, these systems have rich properties of distinctive IR geometries \cite{Ren:2019lgw, Ren:2021rhx}. Thus, generalizing these hairy black holes with an acceleration is expected to have more fruitful applications. Like the original C metric, the charged dilaton C metrics also have a conical singularity along one or both polar axes. This conical singularity can be replaced by a magnetic flux tube \cite{Dowker:1993bt} or a cosmic string (topological defect) \cite{Gregory:1995hd}, giving a precise explanation of the mechanism of the acceleration. Remarkably, after uplifting to $D=11$ supergravity, this conical singularity could be completely removed by imposing proper restrictions on the parameters \cite{Ferrero:2020twa}. Furthermore, these solutions can also be generalized to multiply charged accelerating black holes \cite{Lu:2014sza}. There are also the nonspinning spindles from multicharge accelerating black holes \cite{Ferrero:2021ovq}.

In the context of the AdS/CFT correspondence, supersymmetry also plays a very important role in black hole thermodynamics and holographic interpretation of the uplifting solutions. A significant progress for black hole thermodynamics in recent years is that the Bekenstein-Hawking entropy of supersymmetric AdS solutions can be calculated from their dual supersymmetric conformal field theories (SCFTs) \cite{Benini:2015eyy, Cabo-Bizet:2018ehj}. There are supersymmetric solutions from magnetically charged and accelerating black holes with proper choice of physical parameters \cite{Ferrero:2020twa, Cassani:2021dwa, Ferrero:2021ovq}. Special case of the solutions we analyze are also expected to be supersymmetric under certain conditions. Therefore, it could be very interesting to calculate the entropy of these solutions from their dual SCFTs living on the curved conformal boundary background and compare it with other methods. However, in general, it is difficult to define such a statistical ensemble of the dual SCFTs on a curved background with conical deficits to compute physical quantities. Fortunately, we can calculate the corresponding quantities on the gravity side, which is the goal of this paper.

Here we consider the thermodynamics of slowly accelerating black holes in asymptotically AdS spacetime \cite{Podolsky:2002nk} with scalar hair and electric charge. Since there are no accelerating horizons, the thermodynamics of these systems are clear. If a scalar field exists, however, then the thermodynamics will become more fascinating. On the one hand, it was proposed that the first law of hairy black hole thermodynamics will be modified with a ``scalar charge'' \cite{Gibbons:1996af,Lu:2013ura,Lu:2014maa}. On the other hand, these papers \cite{Caldarelli:2016nni,Astefanesei:2018vga} show that properly choosing boundary terms associated with more general boundary conditions of the scalar field leads to unmodified thermodynamics. It will be very interesting to analyze the hairy C metrics along this line. See section 3 for details. Without introducing a scalar charge, we drive the first law of thermodynamics \cite{Appels:2017xoe,Cassani:2021dwa} and the Smarr relation \cite{Smarr:1972kt, Kastor:2009wy} for hairy slowly accelerating black holes as follows:
\begin{align}
\delta M &= T \delta S + V \delta P + \sum\limits_{I=1}\limits^{2}\Phi_I\delta Q_I - \lambda_+ \delta \mu_+-\lambda_- \delta \mu_--\eta\frac{\delta G_4}{G_4}\label{first},\\
M &= 2 \left( T S- P V \right) + \sum\limits_{I=1}\limits^{2}\Phi_I Q_I,\label{Smarr}
\end{align}
where
\begin{align}
\eta=M-TS-\sum\limits_{I=1}\limits^{2}\Phi_I Q_I-PV+\lambda_+\mu_++\lambda_-\mu_-.
\end{align}
If we employ the duality relation as in \cite{Cong:2021fnf}
\begin{align}
C=k\frac{\ell^2}{16\pi G_4},
\end{align}
after some simple algebra, we can rewrite the first law~\eqref{first} as follows:
\begin{align}
\delta M&=T\delta S+\sum\limits_{I=1}\limits^{2}\Phi_I\delta Q_I+V_c\delta P+\mu_c\delta C-\lambda_+\delta \mu_+-\lambda_- \delta \mu_-\label{first2},
\end{align}
where
\begin{align}
V_c=V+\frac{\eta}{2P}\,,\qquad \mu_c=\frac{\eta}{2C}.
\end{align}
This generalizes the result of \cite{Cong:2021fnf} with conical defects in the bulk spacetime.

This paper is organized as follows. In section $2$, we introduce the solutions of charged dilaton accelerating black holes. In section $3$, we take a brief review on boundary counterterms, boundary conditions for the scalar field, and the holographic dictionary, then obtain the holographic stress-energy tensors of these solutions to compute the holographic mass. To keep conformal symmetry, we also propose a new boundary condition. In section $4$, we derive the standard thermodynamic laws for slowly accelerating electrically charged dilaton black holes. In section $5$, we give the thermodynamics of $U(1)^2$-charged accelerating black holes. In section $6$, we summarize our results and discuss generalizations to black funnels and droplets \cite{Hubeny:2009kz, Caldarelli:2011wa}. In the Appendix A, we give some details regarding the special cases of gauged supergravity.

\section{Accelerating black holes in Einstein-Maxwell-dilaton systems}

Before discussing the black hole thermodynamics, we introduce the solutions of accelerating charged dilaton black holes, which can be found in \cite{Lu:2014ida, Ren:2019lgw}. The bulk action is
\begin{align}
S_{\text{bulk}}&=\frac{1}{16 \pi G_4}\int_{\mathcal M}\!d^{3+1}x \sqrt{-G}\left(R-\frac{1}{2}(\partial\phi)^2-V(\phi)-\frac{1}{4}Z(\phi)\mathcal F^2\right),
\label{eq:Sbulk}
\end{align}
where $\mathcal F=dA$, $Z(\phi)=e^{-\zeta \phi}$, and the scalar potential is \cite{Gao:2004tu}
\begin{align}\label{VV}
V(\phi)=-\frac{2e^{-\frac{\phi}{\zeta}}\left(\zeta^2(3\zeta^2-1)+8\zeta^2e^{\frac{1+\zeta^2}{2\zeta}\phi}+(3-\zeta^2)e^{\zeta\phi+\frac{\phi}{\zeta}}\right)}{(1+\zeta ^2)^2\ell^2},
\end{align}
which can also be expressed with a superpotential
\begin{align}
V = \left(\frac{dW}{d\phi}\right)^2-\frac{3}{4}W^2\,,\qquad
W = \frac{\sqrt{8}}{(1+\zeta^2)\ell}\left(e^{\frac{1}{2}\zeta\phi}+\zeta^2e^{-\frac{1}{2\zeta}\phi}\right).
\end{align}
Here $\zeta$ is the dilaton coupling constant, and $\ell$ is the AdS radius. When $\zeta=0$, $1/\sqrt{3}$, $1$, and $\sqrt{3}$, these solutions can be embedded in gauged supergravities (see Appendix A for more details). In particular, the $\zeta=0$ case corresponds to Reissner–Nordström-AdS (RN-AdS) accelerating black hole, which has been explored before, and the goal of this paper is to analyze the other solutions. The equations of motion are invariant under an electric-magnetic duality transformation and $\phi \to-\phi$ \cite{Dowker:1993bt}, and the discussion of the magnetic-charged dilation black holes is similar to electric-charged ones. Therefore, we will only focus on electrically charged solutions.

The line elements describing accelerating black holes with scalar hair and electric charge can be written as follows:
\begin{equation}
ds^2=\frac{1}{a^2(x-y)^2}\biggl[f_x^{\frac{2 \zeta^2}{1+\zeta^2}}\biggl(-\frac{F_y}{\alpha^2}dt'^2+\frac{dy^2}{F_y}\biggr)+h_y^{\frac{2 \zeta^2}{1+\zeta^2}}\biggl(\frac{dx^2}{G_x}+\frac{G_x}{K^2}d\varphi^2\biggr)\biggr],
\end{equation}
with
\begin{align}
&F_y=-(1-y^2)(1+2acy)h_y^{\frac{1-\zeta^2}{1+\zeta^2}}+\frac{1}{a^2\ell^2}h_y^{\frac{2\zeta^2}{1+\zeta^2}}\,,\nonumber\\
&G_x=(1-x^2)(1+2acx)f_x^{\frac{1-\zeta^2}{1+\zeta^2}}\,,\nonumber\\
&h_y=1+aby\,,\qquad f_x=1+abx\,,\\
&A=\frac{2}{\alpha}\sqrt{\frac{2bc}{1+\zeta^2}}(y-y_h)dt'\,,\qquad e^{\zeta\phi}=\left(\frac{h_y}{f_x}\right)^{\frac{2\zeta^2}{1+\zeta^2}}.\nonumber
\end{align}
The AdS boundary of these solutions are at $x=y$, and the boundary metrics are
\begin{equation}
ds_\text{bdy}^2=f_x^{\frac{2 \zeta^2}{1+\zeta^2}}\biggl(-\frac{F_x}{\alpha^2}dt'^2+\frac{f_x^{\frac{2 \zeta^2}{1+\zeta^2}}}{a^2\ell^2F_xG_x}dx^2+\frac{G_x}{K^2}d\varphi^2\biggr).
\end{equation}
The coordinate transformation $y=-\frac{1}{ar}$, $x=\cos(\theta)$, $t'=at$ takes the metrics to another form
\begin{equation}
ds^2=\frac{1}{H(r,\theta)^2}\biggl[f(\theta)^{\frac{2\zeta^2}{1+\zeta^2}}\biggl(-\frac{F(r)}{\alpha^2}dt^2+\frac{dr^2}{F(r)}\biggr)+r^2h(r)^{\frac{2\zeta^2}{1+\zeta^2}}\biggl(\frac{d\theta^2}{G(\theta)}+\frac{\sin^2\theta G(\theta)d\varphi^2}{K^2}\biggr)\biggr],
\end{equation}
with
\begin{align}
&H=1+ar\cos\theta\,,\nonumber\\
&F=(1-a^2r^2)\Big(1-\frac{2c}{r}\Big)h^{\frac{1-\zeta^2}{1+\zeta^2}}+\frac{r^2}{\ell^2}h^{\frac{2\zeta^2}{1+\zeta^2}}\,,\nonumber\\
&G=(1+2ac\cos\theta)f^{\frac{1-\zeta^2}{1+\zeta^2}}\,,\nonumber\\
&h=1-\frac{b}{r}\,,\qquad f=1+ab\cos\theta\,,\\
&A=\frac{2}{\alpha}\sqrt{\frac{2bc}{1+\zeta^2}}\left(\frac{1}{r_h}-\frac{1}{r}\right)dt\,,\qquad e^{\zeta\phi}=\left(\frac{h}{f}\right)^{\frac{2\zeta^2}{1+\zeta^2}}.\nonumber
\end{align}

The parameters appearing in the metrics $c$ and $a$ are usual parameters related to the mass and acceleration of the black holes. While the scalar field is associated with the parameter $b$, the parameters $b$ and $c$ determine the gauge potential together. When $b$ vanishes, the metrics reduce to usual accelerating black hole in AdS \cite{Plebanski:1976gy, Dias:2002mi} without any charges. In the limit $\ell\to\infty$, up to a scalar field rescaling, these solutions recover accelerating dilaton black holes in string theory. In particular, when $\zeta\textless1$ these solutions correspond to pair creation of dilaton black holes \cite{Dowker:1993bt}. And another limit $a\to 0$ recovers usual dilaton black holes in asymptotically AdS spacetime \cite{Gao:2004tu}.

Analogous to previous works \cite{Anabalon:2018ydc,Anabalon:2018qfv,Cassani:2021dwa}, to obtain correct thermodynamic quantities for slowly accelerating charged dilaton black holes in asymptotically AdS spacetime, it is vital to rescale the time coordinate with $\alpha$. Only with a correct choice of $\alpha$ (which is not unique) can we consistently verify the first law of thermodynamics for accelerating black holes. The parameter $K$ carries the information of conical deficit of these spacetimes at two axes, and then it gives the tension defined by \cite{Appels:2017xoe,Cassani:2021dwa} (where $\theta_+=0$, $\theta_-=\pi$)
\begin{equation}\label{mupm}
\mu_{\pm}=\frac{\delta_\pm}{8\pi G_4}=\frac{1}{4G_4}\biggl(1-\frac{G(\theta_\pm)}{K}\biggr)=\frac{1}{4G_4}\biggl(1-\frac{(1\pm 2ac)(1\pm ab)^{\frac{1-\zeta^2}{1+\zeta^2}}}{K}\biggr).
\end{equation}

Finally, to maintain the Lorentz signature for the metrics describing slowly accelerating black holes with scalar hair and electric charge, we need to constrain the range of parameters as follows:
\begin{itemize}
\item It is obvious that the conformal boundary of these spacetime are located at $r_\text{bdy}=-1/a\cos\theta$. To obtain slowly accelerating solutions (no acceleration horizon), we must have $a\cos\theta\le1/r\le1/r_h$ and $F(r_\text{bdy})>0$ \cite{Podolsky:2002nk}.

\item For the $\theta$ coordinate, we require $G(\theta)\textgreater0$ and $f(\theta)\textgreater0$ on $\theta\in [0,\pi]$, which lead to $ac\textless\frac{1}{2}$ and $ab\textless1$.

\item The spacetime is singular when $r\to 0$ or $r\to b$, and the later requires $r_h\textgreater b$ coincidence with the restriction on $h(r)\textgreater0$. To have a black hole without naked singularity, we require $F(r)$ have at least one root in the range $r\in(b,1/a)$. 
\end{itemize}

For these ranges of the parameters, these metrics correspond to slowly accelerating black holes with electric charge and scalar hair in asymptotically AdS spacetime. Consequently, we can discuss the thermodynamics of these black holes in equilibrium without ambiguity.

\section{Holographic Renormalization}

As our systems include a scalar field, there are more subtleties. In particular, properly renormalizing the stress-energy tensors with mixed boundary conditions for the scalar field is important for correctly giving the thermodynamic relations of these solutions. Holographic renormalization for charged dilaton black holes with general boundary conditions for the scalar field was studied in detail in \cite{Caldarelli:2016nni}. In this section, we will take a brief review of boundary counterterms, boundary conditions for the scalar field, and the holographic dictionary, then calculate the holographic stress-energy tensors and mass for these solutions. To keep conformal symmetry on the boundary, we also propose a new boundary condition for these solutions we study. 

\subsection{Holographic dictionary for mixed boundary conditions}
To obtain the counterterms, the Fefferman-Graham (FG) \cite{FG} expansion is needed near the AdS boundary. In the FG gauge, the metric takes the following form:
\begin{equation}\label{FGg}
ds^2=\frac{\ell^2}{z^2}\left(dz^2+g_{ij}(x,z)dx^idx^j\right),
\end{equation}
with
\begin{align}
g_{ij}(x,z)&=g(x)_{(0)ij}+zg(x)_{(1)ij}+z^2g(x)_{(2)ij}+\cdots\,,\nonumber\\
A_i(x,z)&=A_i(x)_{(0)}+zA_i(x)_{(1)}+z^2A_i(x)_{(2)}+\cdots\,,\\
\phi(x,z)&=z^{\Delta_-}\varphi(x,z) =z^{\Delta_-}\left(\varphi(x)_{(0)}+z\varphi(x)_{(1)}+z^2\varphi(x)_{(2)}+\cdots\right).\nonumber
\end{align}
Defining the following constants \cite{Caldarelli:2016nni}
\begin{align}
V_k\equiv\lim_{r\rightarrow r_\text{bdy}}V^{(k)}(\phi),\qquad
Z_k\equiv\lim_{r\rightarrow r_\text{bdy}}Z^{(k)}(\phi),
\end{align}
where $V^{(k)}(\phi)$ and $Z^{(k)}(\phi)$ are the $k$th derivative of $V(\phi)$ and $Z(\phi)$, respectively. In our systems, we have $\Delta_-=1$ and the scalar field vanishes at the boundary, so the scalar potential has the asymptotic expansion near the boundary
\begin{align}
V(\phi)=-\frac{6}{\ell^2}-\frac{\phi ^2}{\ell^2}+\frac{1-4 \zeta^2+\zeta^4}{24\zeta^2 \ell^2}\phi ^4+\cdots.
\end{align}
Then the AdS mass of the scalar field is
\begin{align}
m^2_\phi=V_2=-\frac{2}{\ell^2},
\end{align}
which lies in the range
\begin{align}
-\frac{3^2}{4}+1\ge\ell^2 m^2_\phi\ge-\frac{3^2}{4}.
\end{align}
Therefore, we can consider more general (Neumann or mixed) boundary conditions for the scalar field, which can lead to the first law of thermodynamics without any scalar charge \cite{Caldarelli:2016nni}.

In the FG expansion, the equations of motion determine the relation of the coefficients.\footnote{The relation of the coefficients of the FG expansion is
\begin{align}\label{FGCR}
&g(x)_{(1)ij}=0,\nonumber\\
&g(x)_{(2)ij}+R_{(0)ij}-\frac{1}{4}R_{(0)}g(x)_{(0)ij}+\frac{1}{8}\varphi(x)^2_{(0)}g(x)_{(0)ij}=0,\nonumber\\
&g(x)_{(0)}^{jk}\left(D_{(0)j}g(x)_{(2)ki}-D_{(0)i}g(x)_{(2)jk}\right)-\frac{1}{2}\varphi(x)_{(0)}\partial_i\varphi(x)_{(0)}=0,\\
&\varphi(x)_{(2)}+\frac{1}{2}\Box_{(0)}\varphi(x)_{(0)}+\frac{1}{2}\text{Tr}\Big(g^{-1}_{(0)}g_{(2)}\Big)\varphi(x)_{(0)}-\frac{1}{12}\ell^2V_4\varphi(x)^3_{(0)}=0,\nonumber
\end{align}
where $D_{(0)i}$, $\Box_{(0)}$, and $R^{(0)}$ are the covariant derivative, Laplacian, and Ricci scalar with respect to $g(x)_{(0)ij}$.}
As a consequence, these relations lead to covariant boundary counterterms at the cutoff $z=\epsilon$ as follows:
\begin{align}
S_{\text{ct}}=-\frac{1}{16 \pi G_4}\int_{z=\epsilon}\!d^3x\sqrt{-\gamma}\left(\frac{4}{\ell}+\ell R[\gamma]+\frac{1}{2\ell}\phi^2\right),
\end{align}
where $\gamma_{ij}$ is the induced metric on the boundary. With this counterterm, we can holographically renormalize the action for an asymptotically AdS spacetime with the Gibbons-Hawking term
\begin{align}
S_{\text{GH}}=\frac{1}{16 \pi G_4}\int_{z=\epsilon} d^3x \sqrt{-\gamma}\;2 \mathcal K,
\end{align}
where $\mathcal K$ is the trace of extrinsic curvature with respect to the induced metric. Then the renormalized action independent of the cutoff $\epsilon$ is
\begin{align}
S_{\text{ren}}=\lim_{\epsilon\to 0}\left(S_{\text{bulk}}+S_{\text{GH}}+S_{\text{ct}}\right),
\end{align}
where $S_\text{bulk}$ is the action~\eqref{eq:Sbulk} with $r$ being integrated to $z=\epsilon$.

Since the scalar field admits more general boundary conditions, the choice of the source \cite{Caldarelli:2016nni}
\begin{equation}
J_{\mathcal F_\varphi}=-\ell^2\varphi_{(1)}-\mathcal F'_\varphi(\varphi_{(0)}),
\end{equation}
where $\mathcal F_\varphi(\varphi_{(0)})=\frac{1}{3}x_1\varphi^3_{(0)}$, corresponds to a triple-trace deformation in the dual CFT \cite{Witten:2001ua, Mueck:2002gm,Papadimitriou:2007sj}. Specifying the value of $x_1$ is important for correctly giving the laws of thermodynamics for these solutions, and this choice also requires an additional term for the on-shell action as follows:
\begin{align}\label{SFF}
S_{\mathcal F_\varphi}=\frac{1}{16\pi G_4}\int_{z=\epsilon}d^3x\sqrt{-g_{(0)}}\Big(J_{\mathcal F_\varphi}\varphi_{(0)}+\mathcal F_\varphi(\varphi_{(0)})\Big). 
\end{align}
Then the modified one-point functions are (T denotes triple-trace deformation)
\begin{align}\label{TmumuT}
\langle \mathcal J_i\rangle_\text{T}&=\frac{1}{16\pi G_4}Z_0A_{i(1)},\nonumber\\
\langle\mathcal O_{\Delta_-}\rangle_\text{T}&=\frac{1}{16 \pi G_4}\varphi_{(0)},\nonumber\\
\langle \mathcal T_{ij}\rangle_\text{T}&=\frac{1}{16 \pi G_4}\Big[3\ell^2g_{(3)ij}+\Big(\mathcal F_\varphi(\varphi_{(0)})-\varphi_{(0)}\mathcal F'_\varphi(\varphi_{(0)})\Big)g_{(0)ij}\Big],\\
&=\frac{\ell^2}{16\pi G_4}\Big(3g_{(3)ij}-\frac{2}{3}\frac{x_1}{\ell^2}\varphi_{(0)}^3g_{(0)ij}\Big),\nonumber
\end{align}
and the local Ward identities are
\begin{align}\label{WDIT}
D_{(0)i}\langle \mathcal J^i\rangle_\text{T}&=0,\nonumber\\
D_{(0)}^j\langle \mathcal T_{ij}\rangle_\text{T}&=\langle\mathcal O_{\Delta_-}\rangle_\text{T}\partial_i J_{\mathcal F_\varphi}+\langle\mathcal J^j\rangle_\text{T} \mathcal F_{ij}^{(0)},\\
\langle \mathcal T^i{}_i\rangle_\text{T}&=\frac{1}{16\pi G_4}\Big[2\varphi_{(0)}J_{\mathcal F_\varphi}+3\mathcal F_\varphi(\varphi_{(0)})-\varphi_{(0)}\mathcal F'_\varphi(\varphi_{(0)})\Big].\nonumber
\end{align}

Although these solutions admit a triple-trace deformation (the $\zeta=1$ solution also admits Dirichlet boundary conditions), the source $J_{\mathcal F_\varphi}$ and the trace of the $\langle \mathcal T_{ij}\rangle_\text{T}$ of these solutions we study are nonzero. We propose a new boundary condition that preserves the conformal symmetry at the AdS boundary, and we call it conformal boundary condition. We solve the value of $x_1$ from $J_{\mathcal F_\varphi}=0$, and we obtain $x_1=-\ell^2\varphi_{(1)}/\varphi_{(0)}^2$. Since the value of $-\ell^2\varphi_{(1)}/\varphi_{(0)}^2$ is coordinate dependent for accelerating black holes, this choice of $x_1$ leads to a different boundary condition and these two kinds of boundary conditions reduce to the same one when $-\ell^2\varphi_{(1)}/\varphi_{(0)}^2$ is coordinate independent. We can put this $x_1$ in~\eqref{SFF} and~\eqref{TmumuT} to obtain the additional term for the on-shell action and the one-point function. However, the local Ward identities~\eqref{WDIT} are modified for the conformal boundary condition.

The additional term for the on-shell action, the stress-energy tensor and the local Ward identities\footnote{The local Ward identity~\eqref{TUN11} is obtained by putting \eqref{TUN} in (A.18) of \cite{Caldarelli:2016nni}.} for the conformal boundary condition are (C denotes conformal)
\begin{align}
S_{\text{C}}&=\frac{1}{16\pi G_4}\int_{z=\epsilon}d^3x\sqrt{-g_{(0)}}\Big(-\frac{\ell^2}{3}\varphi_{(1)}\varphi_{(0)}\Big),\\
\langle \mathcal T_{ij}\rangle_\text{C}&=\frac{\ell^2}{16\pi G_4}\Big(3g_{(3)ij}+\frac{2}{3}\varphi_{(0)}\varphi_{(1)}g_{(0)ij}\Big),\label{TUN}\\
D_{(0)}^j\langle \mathcal T_{ij}\rangle_\text{C}&=\frac{\ell^2}{16\pi G_4}\Big(\frac{2}{3}\varphi_{(1)}\partial_i\varphi_{(0)}-\frac{1}{3}\varphi_{(0)}\partial_i\varphi_{(1)}\Big)+\langle\mathcal J^j\rangle_\text{T} \mathcal F_{ij}^{(0)},\label{TUN11}\\
\langle \mathcal T^i{}_i\rangle_\text{C}&=0.
\end{align}
Since the stress-energy tensor is traceless, the conformal symmetry is preserved. We are going to see that our solutions fit perfectly in the framework of either boundary condition in the next subsection.

\subsection{Holographic mass of accelerating charged dilaton black holes}

With these techniques, we calculate the holographic mass from the boundary stress-energy tensors. First, it is necessary to use the FG coordinates, and we take the same coordinate transformation as in \cite{Anabalon:2018qfv}
\begin{equation}\label{FGT}
\frac{1}{r}= -a\rho-\sum_{n=1}^4F_{n}(\rho)z^{n},\; \quad \cos \theta = \rho+\sum_{n=1}^4G_{n}(\rho) z^{n}.\\
\end{equation}
The functions $F_n(\rho)$ and $G_n(\rho)$ can be solved by requiring $g_{zz}\equiv\ell^2/z^2$ and $g_{z\rho}\equiv0$ at each order of $z$ in the FG gauge~\eqref{FGg}, and the leading order gives
\begin{equation}
G_{1}(\rho)=-\frac{a\ell^2 F_{1}(\rho)X(\rho)}{\gamma_b(\rho)^2},\\
\end{equation}
where
\begin{align}
X(\rho )&=(1-\rho^2)(1+2ac\rho ),\\
\gamma_b(\rho )&=\sqrt{(1+ab\rho)^{\frac{3\zeta^2-1}{1+\zeta^2}} -a^2 \ell^2X(\rho )}.
\end{align}
Note that the square root requires $\gamma^2_b(\rho )\textgreater 0$, which is the same as $F(r_\text{bdy})\textgreater0$. To simplify calculations, we choose $F_{1}(\rho )=-\gamma^2_b(\rho )/\alpha$, and then the boundary metric can be written as
\begin{equation}\label{mybdmetric}
ds^2_{(0)}=-\frac{(1+ab\rho)^{\frac{3-5\zeta^2}{1+\zeta^2}}\gamma^2_b(\rho )}{\ell^4}dt^2+\frac{\alpha ^2}{\ell^2 X(\rho ) \gamma _b(\rho )^2}d\rho^2+\frac{(1+ab\rho)^{\frac{3-5\zeta^2}{1+\zeta^2}}\alpha ^2 X(\rho )}{K^2 \ell^2}d\varphi^2.
\end{equation}

The electric-magnetic duality transformation does not change the bulk metric, so the boundary metric of the magnetically charged solution is the same as~\eqref{mybdmetric}. As pointed out in \cite{Ferrero:2020twa, Ferrero:2021ovq}, the boundary metric of the magnetically charged solution can describe a spindle, warped by the $d=3$ SCFTs on the M2-branes. After uplifting to $D=11$ supergravity, the solution can be completely regular with proper parameters choice. An interesting feature is that warped M2-brane theories in the UV flow to a $d=1$ supersymmetric conformal quantum mechanics in the IR, which is dual to the near horizon limit AdS$_2\times Y_9$ solution of the black holes. In particular, the $\zeta=1$ case has been studied in \cite{Ferrero:2021ovq}, and the others need further study in future work.

Since the scalar field admits more general boundary conditions in our systems, we take the mixed boundary conditions for these solutions. For the triple-trace boundary condition, we choose the value of $x_1$ as
\begin{align}\label{x11}
x_1=\frac{1-\zeta^2}{4\zeta}\ell^2, 
\end{align}
by requiring that the holographic mass obtained by the two boundary conditions discussed above are the same, and we will show that it is also the same as the Ashtekar-Magnon-Das (AMD) mass~\eqref{eq:AMD} below. This choice of $x_1$ is consistent with the 
nonaccelerating black holes of these systems \cite{Ren:2019lgw} with standard thermodynamic laws. And we can find that the triple-trace and conformal boundary conditions reduce to the same one for nonaccelerated black holes. As a result, the expectation value of the stress-energy tensor~\eqref{TmumuT} is
\begin{equation}\label{myTmumu13}
\langle\mathcal T_{i}^{j}\rangle_\text{T}=\langle\mathcal T_{i}^{j}\rangle_\text{C}+\frac{a\ell^4\zeta^2\omega_bb^2(1+ab\rho)^{\frac{3\zeta^2-5}{1+\zeta^2}}}{12\pi G_4\alpha^3(1+\zeta^2)^2}\text{diag}(1,1,1),
\end{equation}
and the expectation value of the stress-energy tensor~\eqref{TUN} is
\begin{equation}\label{TUNC}
\langle\mathcal T_{i}^{j}\rangle_\text{C}=\frac{\ell^4(1+ab\rho)^{\frac{-4}{1+\zeta^2}}\omega_E}{48\pi G_4\alpha^3(1+\zeta^2)^2}\text{diag}\left[-\left(2\gamma_b^2-a^2\ell^2X\right),(1+ab\rho)^{\frac{3\zeta^2-1}{1+\zeta^2}},\gamma_b^2-2a^2\ell^2X\right],
\end{equation}
where
\begin{align}
\omega_1&=1-\zeta^2+12ac\rho,\nonumber\\
\omega_2&=(3-\zeta^2)\rho+ac\left(27\rho^2-3\zeta^4+\zeta^2(1+15\rho^2)\right),\nonumber\\
\omega_3&=3\zeta^4+3\rho^2(1+8ac\rho)-\zeta^2\left(1+\rho^2(1-8ac\rho)\right),\\
\omega_b&=2\rho+a\left(b-2c+(b+6c)\rho^2+4abc\rho^3\right),\nonumber\\
\omega_E&=6c(1+\zeta^2)^2+3b(1+\zeta^2)\omega_1+2ab^2\omega_2+a^2b^3\omega_3.\nonumber
\end{align}
With these results, we can quickly check that the stress-energy tensor~\eqref{myTmumu13} is not traceless, but the stress-energy tensor~\eqref{TUNC} is traceless, which is expected for the conformal boundary condition. When $b=0$, in which case the scalar field vanishes, the boundary metric~\eqref{mybdmetric} and the stress-energy tensors~\eqref{myTmumu13} and \eqref{TUNC} reduce to the results in \cite{Anabalon:2018qfv} consistently.

With the stress-energy tensors in hand, we obtain the holographic mass for these solutions by integrating the spatial dimensions:
\begin{align}\label{massintegrate}
M&=\int{-\langle\mathcal T_{t}^{t}\rangle_\text{T}\sqrt{-g_{(0)}}\;d\rho d\varphi}=\int{-\langle\mathcal T_{t}^{t}\rangle_\text{C}\sqrt{-g_{(0)}}\;d\rho d\varphi}\nonumber\\
&=\frac{c+\frac{1-\zeta^2}{1+\zeta^2}\frac{b}{2}+\frac{1}{8}a\ell^2\Big((1-2ac)^2(1-ab)^{2\frac{1-\zeta^2}{1+\zeta^2}}-(1+2ac)^2(1+ab)^{2\frac{1-\zeta^2}{1+\zeta^2}}\Big)}{\alpha KG_4},
\end{align}
where $ab\textless 1$, which is satisfied with our parameter constraint. It turns out that the last terms in \eqref{myTmumu13} proportional to $\omega_b$ do not contribute to the integral in \eqref{massintegrate} for calculating the mass.

The asymptotic expansions of the gauge potential and the scalar field are
\begin{align}
A_t(\rho,z)&=A_t(\rho)_{(0)}+zA_t(\rho)_{(1)}+z^2A_t(\rho)_{(2)}+\cdots\,,\nonumber\\
\phi(\rho,z)&=z\left(\varphi(\rho)_{(0)}+z\varphi(\rho)_{(1)}+z^2\varphi(\rho)_{(2)}+\cdots\right),
\end{align}
with
\begin{align}
A_t(\rho)_{(0)}=&\frac{2}{\alpha r_h}\sqrt{\frac{2bc}{1+\zeta^2}}(1+ar_h\rho),\nonumber\\
A_t(\rho)_{(1)}&=-\frac{2}{\alpha^2}\sqrt{\frac{2bc}{1+\zeta^2}}\gamma _b(\rho){}^2,\nonumber\\
\varphi(\rho)_{(0)}=&-\frac{2b\zeta(1+ab\rho)^{2\frac{\zeta^2-1}{1+\zeta^2}}}{\alpha(1+\zeta^2)},\\
\varphi(\rho)_{(1)}=&\frac{b\zeta(1+ab\rho)^{\frac{\zeta^2-3}{1+\zeta^2}}}{\alpha^2(1+\zeta^2)}
\Big[a\ell^2\left(2\rho+4a^2bc\rho^3+a(b-2c+b\rho^2+6c\rho^2)\right)\nonumber\\
&+b\frac{\zeta^2-1}{1+\zeta^2}(1+ab\rho)^{\frac{3\zeta^2-1}{1+\zeta^2}}\Big].\nonumber
\end{align}
Then we can easily get the dual field theory one-point functions~\eqref{TmumuT} and check the local Ward identities~\eqref{WDIT} and \eqref{TUN11} with the stress-energy tensors~\eqref{myTmumu13} and~\eqref{TUNC} respectively. With $\varphi(\rho)_{(2)}$, $g(\rho)_{(1)ij}$, and $g(\rho)_{(2)ij}$, we can also check~\eqref{FGCR}, but it is not illuminating to write everything down.

\section{Thermodynamics of slowly accelerating charged dilaton black holes}

We begin this section by writing down thermodynamic quantities first:
\begin{align}\label{THQ}
M&=\frac{c+a_1\frac{b}{2}+\frac{1}{8}a\ell^2\Big((1-2ac)^2(1-ab)^{2a_1}-(1+2ac)^2(1+ab)^{2a_1}\Big)}{\alpha KG_4},\nonumber\\
\alpha&=\sqrt{a\alpha KG_4M},\qquad S=\frac{\pi r_h^2h(r_h)^{1-a_1}}{KG_4\left(1-a^2r_h^2\right)},\nonumber\\
T&=\frac{h(r_h)^{a_1}}{4\pi\alpha r_h^2}\biggl[2c(1-a^2r_h^2)+\frac{r_h^2\Big(2r_h-b\left(1+2a_1+(1-2a_1)a^2r_h^2\right)\Big)}{\ell^2h(r_h)^{2a_1}(1-a^2r_h^2)}\biggr],\nonumber\\
Q&=\frac{\sqrt{(1+a_1)bc}}{2KG_4}, \qquad \Phi=\frac{2\sqrt{(1+a_1)bc}}{\alpha r_h},\qquad P=\frac{3}{8\pi\ell^2G_4},\\
V&=\frac{4\pi\ell^2}{3\alpha K}\Big[c\Big(1+\frac{a_1b}{r_h}\Big)-\alpha KMG_4+\frac{r_h^2\Bigl(2r_h-b\left(1+2a_1+(1-2a_1)a^2r_h^2\right)\Bigr)}{2\ell^2h(r_h)^{2a_1-1}(1-a^2r_h^2)^2}\Big],\nonumber\\
\lambda_\pm&=\frac{1}{\alpha L_1}\Big[L_2\left(L_3L_{3\pm}-L_4L_{4\pm}+L_5L_{5\pm}\right)-\left(L_6L_{3\pm}+L_7L_{5\pm}+L_8L_{8\pm}\right)L_9\Big],\nonumber
\end{align}
with
\begin{align}\label{THLG}
&a_1=\frac{1-\zeta^2}{1+\zeta^2},\quad h(r_h)=1-\frac{b}{r_h},\quad L_{1\pm}=a_1^2b^2(1\mp2ac),\quad L_{2\pm}=a_1b(1\mp ab)(2ac\mp1),\nonumber\\
&L_1=4a\ell^2r_h(1-a^2r_h^2)^3(1-a^2b^2)^{a_1}\left[b^2\left(a_1^2-4a^2(a_1^2-1)c^2\right)-4c^2\right],\nonumber\\
&L_2=(1-a^2r_h^2)h(r_h)^{-2a_1}\left(a^2br_h^2(a_1-1)-(a_1+1)b+2r_h\right),\nonumber\\
&L_3=2a^2\ell^2r_h^2(2c-r_h)h(r_h)^{2a_1},\quad L_4=2\ell^2c(1-a^2r_h^2)h(r_h)^{2a_1},\nonumber\\
&L_5=(a_1-1)br_h^2-a_1b\ell^2(1-a^2r_h^2)(1-\frac{2c}{r_h})h(r_h)^{2a_1-1},\quad L_6=2a^2r_h^3h(r_h),\nonumber\\
&L_7=(a_1-1)b(1-a^2r_h^2),\quad L_8=2ar_h(1-a^2r_h^2)h(r_h),\nonumber\\
&L_{6\pm}=(1\pm ab)(1\mp ab)^{a_1},\quad L_{7\pm}=a^2\ell^2(2ac\pm1)(1\pm ab)^{a_1}(1-a^2b^2)^{a_1},\\
&L_9=2\ell^2c(1-a^2r_h^2)^2+r_h^2\left[2r_h-b\left(a^2(1-2a_1)r_h^2+2a_1+1\right)\right]h(r_h)^{-2a_1},\nonumber\\
&L_{3\pm}=L_{6\pm}\left[2aL_{1\pm}-L_{2\pm}+2c(1\mp ab)(2ac\pm1)\left(a^2\ell^2(1\pm ab)^{2a_1}-1\right)\right]\nonumber\\
&\hspace{0.05\textwidth}+a_1bL_{7\pm}\left(4a^2c(b-c)-1\right),\nonumber\\
&L_{4\pm}=L_{6\pm}\left[L_{2\pm}-2aL_{1\pm}+2c(ab\mp1)\left(1\mp6ac+a^2\ell^2(1\pm2ac)(1\pm ab)^{2a_1}\right)\right]\nonumber\\
&\hspace{0.05\textwidth}-a_1bL_{7\pm}\left(4a^2c(b-3c)+1\right),\nonumber\\
&L_{5\pm}=-L_{6\pm}\left(2aL_{1\pm}+L_{2\pm}+2c(ab\mp1)(1\pm2ac)\right)\nonumber\\
&\hspace{0.05\textwidth}+L_{7\pm}\left[2c+b\left(4a^2a_1c(c-b)-2a^2bc+a_1\right)\right],\nonumber\\
&L_{8\pm}=L_{6\pm}\left(L_{1\pm}-4c^2(1\mp ab)\right)+2abca_1(b-2c)L_{7\pm}.\nonumber
\end{align}
Consequently, together with the tensions $\mu_{\pm}$ defined by~\eqref{mupm} we can quickly check that both the first law~\eqref{first} or~\eqref{first2} and Smarr relation~\eqref{Smarr} are satisfied for arbitrary $\zeta$.

The value of $\alpha$ for arbitrary $\zeta$ cannot be directly calculated. However, when $\zeta=1$, $1/\sqrt{3}$, $\sqrt{3}$ (cases that can be embedded in gauged supergravities), we find that the corresponding $\alpha$ can be obtained by \eqref{HLMD} below with different stress-energy tensors that are related to different boundary conditions for the scalar field. Thus we write the $\alpha$ and its corresponding $\lambda_\pm$ of these three cases down as follows and discuss an issue of good/bad limit related to them:
\begin{itemize}
\item $\zeta=1$

The $\alpha$ in~\eqref{THQ} and its corresponding $\lambda_\pm$ are
\begin{align}\label{THQ12}
\alpha&=\sqrt{ac(1-a^2\ell^2)},\nonumber\\
\lambda_\pm&=\frac{1}{2\alpha}\left(\frac{\mp1-ab(1\mp2ar_h)}{a(1-a^2r_h^2)}+\frac{2r_h\mp a\left(\ell^2 (1\pm ar_h)^2-r_h^2\right)}{1\pm ar_h}\right).
\end{align}
This $\alpha$ has a bad limit: $\alpha\to 0$ as $a\to 0$. A better $\alpha$ that exists a good limit ($a\to0$, $\alpha\to1$) and its corresponding $\lambda_\pm$ are
\begin{align}\label{THQ1}
\alpha&=\sqrt{1-a^2\ell^2},\qquad\lambda_\pm=\frac{1}{\alpha}\left(\frac{r_h}{1\pm ar_h}-c\mp a\ell^2-\frac{b}{2}\frac{1\mp ar_h}{1\pm ar_h}\right).
\end{align}
Although the $\alpha$ in~\eqref{THQ} does not have a good limit (see more details later), it is highly nontrivial to formulate the first law. The nonuniqueness of $\alpha$ is related to the nonuniqueness of the solution of partial differential equations (PDEs) below we solve. Different $\alpha$ leads to different $\lambda_\pm$. In particular, we can obtain the better $\alpha$ only for the $\zeta=1$ solution.

\item $\zeta=1/\sqrt{3}$
\begin{align}\label{THQ13}
\alpha&=\sqrt{a\Big(c+\frac{b}{4}\Big)\Big(1-a^2\ell^2(1+\frac{a^2bc^2}{c+\frac{b}{4}})\Big)},\nonumber\\
\lambda_\pm&=\frac{\sqrt{1\pm ab}}{\alpha}\left(\frac{2ar_h\mp 1}{2a(1-a^2r_h^2)}-a^2\ell^2c(1\pm \frac{1}{ar_h})\right).
\end{align}

\item $\zeta=\sqrt{3}$
\begin{align}\label{THQ3}
\alpha&=\sqrt{\frac{a\Big(c-\frac{b}{4}\Big)\Big(1-a^2\ell^2(1+\frac{b^2}{\ell^2}-\frac{a^2bc^2}{c-\frac{b}{4}})\Big)}{1-a^2b^2}},\\
\lambda_\pm&=\frac{1}{\alpha\sqrt{1\pm ab}}\left(\frac{(r_h-b)(1-a^2br_h)}{1-a^2r_h^2}-a^2\ell^2c(1\pm \frac{1}{ar_h})\mp \frac{1+a^2br_h(a^2b r_h-2)}{2a(1-a^2r_h^2)}\right).\nonumber
\end{align}
\end{itemize}
Now, let us turn towards how to obtain all of these thermodynamic quantities.
\begin{itemize}
\item $\alpha$:

When $\zeta=1$, $1/\sqrt{3}$, and $\sqrt{3}$, we can use the holographic method \cite{Anabalon:2018ydc, Anabalon:2018qfv} to specify $\alpha$. We write
\begin{align}
\delta g_{(0)}^{ij}&=\frac{\delta g_{(0)}^{ij}}{\delta K}\delta K+\frac{\delta g_{(0)}^{ij}}{\delta a}\delta a+\frac{\delta g_{(0)}^{ij}}{\delta b}\delta b+\frac{\delta g_{(0)}^{ij}}{\delta c}\delta c,\\
\delta \mu_{\pm}&=\frac{\delta \mu_{\pm}}{\delta K}\delta K+\frac{\delta \mu_{\pm}}{\delta a}\delta a+\frac{\delta \mu_{\pm}}{\delta b}\delta b+\frac{\delta \mu_{\pm}}{\delta c}\delta c,
\end{align}
keep $\delta\ell=0$, $\delta G_4=0$, $\delta \mu_{\pm}=0$, and calculate
\begin{align}\label{HLMD}
\delta S=\int_{\partial\mathcal M}\!d^{3}x \sqrt{-g_{(0)}}\langle \mathcal T_{ij}\rangle\delta g_{(0)}^{ij}.
\end{align}
Then we can require $\delta S=0$ to solve PDEs to obtain $\alpha$ as a function of $a$, $b$, $c$, and $\ell$, and the results are~\eqref{THQ12},~\eqref{THQ1},~\eqref{THQ13}, and~\eqref{THQ3}. Since the $\alpha$ in~\eqref{THQ12},~\eqref{THQ13}, and~\eqref{THQ3} can be uniformly written into~\eqref{THQ}, we can see that these three cases are sufficient for us to guess an $\alpha$ for arbitrary $\zeta$.

Now let us give more details about how we obtain the $\alpha$ for these three cases. Since we obtain different stress-energy tensors, we can put any of them in~\eqref{HLMD} to calculate $\alpha$. However, if we would like to consistently obtain the $\alpha$, we need to choose a certain stress-energy tensor related to a specific boundary condition for each solution. For example, for the $\zeta=1$ solution, we choose the stress-energy tensor related to Dirichlet boundary conditions. Since the $\zeta=1$ case admits Dirichlet boundary conditions, which means that this stress-energy tensor can also yield the same mass as other stress-energy tensors related to triple-trace or conformal boundary conditions, we can use it to compute $\alpha$. The corresponding conditions for different solutions are as follows (D denotes Dirichlet boundary conditions, for more details about the Dirichlet boundary conditions see \cite{Caldarelli:2016nni}):
\begin{align}
\zeta&=1\quad\Rightarrow\quad\langle \mathcal T_{ij}\rangle_\text{D}=\frac{\ell^2}{16\pi G_4}\Big(3g_{(3)ij}+\varphi_{(0)}\varphi_{(1)}g_{(0)ij}\Big),\label{D11}\\
\zeta&=\frac{1}{\sqrt{3}}\quad\Rightarrow\quad\langle\mathcal T_{ij}\rangle_\text{T}=\frac{\ell^2}{16\pi G_4}\Big(3g_{(3)ij}-\frac{2}{3}\frac{x_1}{\ell^2}\varphi_{(0)}^3g_{(0)ij}\Big),\\
\zeta&=\sqrt{3}\quad\Rightarrow\quad\langle \mathcal T_{ij}\rangle_\text{C}=\frac{\ell^2}{16\pi G_4}\Big(3g_{(3)ij}+\frac{2}{3}\varphi_{(0)}\varphi_{(1)}g_{(0)ij}\Big).
\end{align}

As an example, if we put~\eqref{D11} in~\eqref{HLMD}, then we can obtain the PDEs of the $\alpha$ for the $\zeta=1$ case as follows:
\begin{align}
&\partial_b\alpha(a,b,c,\ell)=0,\\
&a^2\ell^2\alpha(a,b,c,\ell)+a(1-a^2\ell^2)\partial_a\alpha(a,b,c,\ell)-c(1-a^2\ell^2)\partial_c\alpha(a,b,c,\ell)=0.
\end{align}
The general solution of these PDEs is
\begin{align}
\alpha(a,b,c,\ell)=\alpha_1(ac,\ell)\sqrt{1-a^2\ell^2},
\end{align}
where $\alpha_1$ is a free function, different choices of which lead to the results in~\eqref{THQ12} and~\eqref{THQ1}. The $\zeta=1/\sqrt{3}$ and $\sqrt{3}$ cases have a similar situation.

When $c=b=0$, these solutions reduce to pure AdS in Rindler type coordinate. In order to get pure AdS in global coordinates, it is necessary to take the following coordinate transformation \cite{Podolsky:2002nk}:
\begin{align}
1+\frac{R^2}{\ell^2}=\frac{1+(1-a^2\ell^2)r^2/\ell^2}{(1-a^2\ell^2)H^2}\,,\quad R\sin\vartheta=\frac{r\sin\theta}{H}.
\end{align}
This takes the bulk spacetime to AdS in global coordinates
\begin{equation}
ds^2_\text{AdS}= -\Big(1+\frac{R^2}{\ell^2}\Big)dt^2+\frac{dR^2}{1+\frac{R^2}{\ell^2}}  +R^2\Big(d\vartheta^2+\sin^2\vartheta\frac{d\phi^2}{K^2}\Big)\,,
\end{equation}
where we have chosen $\alpha=\sqrt{1-a^2\ell^2}$. However, when $c=b=0$, our $\alpha$ for arbitrary $\zeta$ in~\eqref{THQ} cannot recover this corresponding value. The better $\alpha$ in~\eqref{THQ1} can recover this corresponding value and the differences are only different $\lambda_\pm$.

In contrast to \cite{Anabalon:2018qfv}, the corresponding better $\alpha=\sqrt{1-a^2\ell^2}$ for the $\zeta=1$ charged dilaton accelerating black hole is independent of the electric charge. As we can see, the $\alpha$ of the RN-AdS accelerating black hole in our parameters is~\eqref{RNALPHA} below, which is charge dependent. The limit $b\to 0$ takes these solutions to Schwarzschild accelerating black hole in AdS. However, except for the better $\alpha$ in~\eqref{THQ1}, this limit cannot take our $\alpha$ in~\eqref{THQ} or~\eqref{THQ12},~\eqref{THQ13},~\eqref{THQ3} to the corresponding results $\alpha=\sqrt{1-a^2\ell^2}$ in \cite{Anabalon:2018ydc, Anabalon:2018qfv}, but the consequence of this difference is only a shift for the thermodynamic lengths ($\lambda_\pm$). The better $\alpha$ for Schwarzschild-AdS accelerating black hole is $\alpha=\sqrt{1-a^2\ell^2}$, and we find that the $\alpha$ which can lead to consistent first law is not unique. We think that there exists better $\alpha$ for arbitrary $\zeta$, which can be reduced to the corresponding results when we take all limits mentioned before. For example, when $\zeta=0$, these solutions reduce to RN-AdS accelerating black hole. Compared with the $\alpha$ in~\eqref{THQ},
\begin{align}
\alpha=\sqrt{a\Big(c+\frac{b}{2}\Big)\Big(1-a^2\ell^2(1+2a^2bc)\Big)},
\end{align}
the corresponding better $\alpha$ found in \cite{Anabalon:2018qfv} in our parameters is
\begin{align}\label{RNALPHA}	
\alpha=\sqrt{\Big(1+2a^2bc\Big)\Big(1-a^2\ell^2(1+2a^2bc)\Big)},
\end{align}
which exists in a good limit. However, for arbitrary $\zeta$, it is not guaranteed to have such limits; a similar situation is that the limit $\ell\to\infty$ cannot take the thermodynamics of accelerating black hole in AdS to flat spacetime (e.g., when $\ell\to\infty$, the $\alpha$ in~\eqref{RNALPHA} does not have a good limit), which is mentioned in \cite{Anabalon:2018qfv}.

\item $M$:

Like \cite{Anabalon:2018ydc, Anabalon:2018qfv}, we compare the holographic mass with the conformal mass by AMD \cite{Ashtekar:1984zz, Ashtekar:1999jx}. First we perform a conformal transformation on the bulk metric, $\bar{g}_{\mu\nu} = {\bar{H}}^2g_{\mu\nu}$, which leads to no divergence near the boundary. Then we can obtain the conserved charge with the integration of a conserved current related to a Killing vector $\xi$:
\begin{equation}
Q(\xi )=\frac{\ell}{8\pi G_4}\lim_{\bar{H} \rightarrow 0}\oint\frac{\ell^{2}}{\bar{H}}N^{\alpha }N^{\beta }\bar{C}^{\nu}{}_{\alpha \mu \beta }\xi _{\nu }d\bar{S}^{\mu },\\
\label{eq:AMD}
\end{equation}
where $\bar{H}=\ell Hr^{-1}$, $N_{\mu }=\partial _{\mu }{\bar{H}}$ is the normal vector of the boundary, $\bar{C}^{\mu }{}_{\alpha \nu \beta }$ is the Weyl tensor of $\bar{g}_{\mu\nu}$, and the tangent surface element of $\bar{H}=0$ is
\begin{equation}
d\bar{S}_{\mu }=\delta^{t}_{\mu }\frac{\ell^{2}f^{\frac{4\zeta^2}{1+\zeta^2}}d(\cos\theta) d\phi}{\alpha K}.
\end{equation}
The conformal mass is given by $M=Q(\partial_t)$, which yields the same result as the holographic method.

\item $T$ and $S$:

To obtain the Hawking temperature of slowly accelerating black holes with charge and scalar hair, we use the Euclidean method as normal:
\begin{equation}
ds^2_{\tau,r}\propto\left[\frac{F(r)}{\alpha^2}d\tau^2+\frac{dr^2}{F(r)}\right] \quad \Rightarrow\quad T=\frac{F'(r_h)}{4\pi\alpha}.
\end{equation}
The entropy is obtained simply by the horizon area formula \cite{Herdeiro:2009vd} as
\begin{equation}
S=\frac{\text{Area}}{4G_4}=\frac{1}{4KG_4}\int \frac{r^2h^{\frac{2\zeta^2}{1+\zeta^2}}\sin\theta}{H^2}d\theta d\phi\Big\vert_{r=r_h}=\frac{\pi r_h^2(1-\frac{b}{r_h})^{\frac{2\zeta^2}{1+\zeta^2}}}{KG_4\left(1-a^2r_h^2\right)}.
\end{equation}

\item $Q$ and $\Phi$:

The electric charge is obtained by Gauss's law integral and the chemical potential is obtained by Hawking-Ross prescription \cite{Hawking:1995ap}
\begin{align}\label{QQ}
Q&=\frac{1}{16\pi G_4}\int_{H=0}\ast \mathcal F=\frac{1}{2KG_4}\sqrt{\frac{2bc}{1+\zeta^2}},\\
\Phi&=\frac{1}{16\pi G_4Q\beta}\int_{\partial\mathcal M}\sqrt{\gamma}n_i\mathcal F^{ij}A_j=\frac{2}{\alpha r_h}\sqrt{\frac{2bc}{1+\zeta^2}},
\end{align}
where $\beta$ is the inverse temperature (periodicity of Euclidean time), $\gamma_{ij}$ is the induced metric, and $n^i$ is the outward pointing unit normal vector of the boundary ($\partial\mathcal M\equiv\{H=0\}$).
\item $V$, $\eta$, and $\lambda_\pm$:

Finally, the condition $F(r_h)=0$ is used to derive the thermodynamic variables $V$, $\eta$, and $\lambda_\pm$. Because the physical quantities $P=3/8\pi\ell^2G_4$ (related to the cosmological constant and Newton's constant) and $\mu_{\pm}$ (tensions of deficit) are fixed, we can utilize the following relation to construct the first law and obtain the remaining quantities:
\begin{align}\label{FRH}
\frac{\partial F_{r_h}}{\partial r_h}\delta r_h+\frac{\partial F_{r_h}}{\partial c}\delta c+\frac{\partial F_{r_h}}{\partial b}\delta b+\frac{\partial F_{r_h}}{\partial \ell}\delta\ell+\frac{\partial F_{r_h}}{\partial a}\delta a+\frac{\partial F_{r_h}}{\partial K}\delta K=0.
\end{align}
After replacing $\delta r_h$ with $\delta S$ and $F'(r_h)$ with $4\pi\alpha T$, fixing the term $T\delta S$, and replacing others with $\delta M$, $\delta Q$, $\delta \mu_\pm$, $\delta G_4$, and $\delta P$, we construct the first law~\eqref{first}. The coefficients related to $\delta P$, $\delta Q$, and $\delta \mu_\pm$ are what we need  in~\eqref{THQ},~\eqref{THLG}, and~\eqref{THQ1}. The coefficient of $\delta G_4$ is correctly $\eta/G_4$ at the same time.

This procedure highly depends on the value of $\alpha$ we input: without a correct $\alpha$, \eqref{FRH} cannot give $\delta M=T\delta S+\cdots$. The Smarr relation can be used as a crosscheck of the thermodynamic quantities we obtained, or alternatively, it can be used to calculate the thermodynamic volume $V$.
\end{itemize}

\section{Thermodynamics of $U(1)^2$-charged accelerating black holes}

Following the analysis of accelerating charged dilaton black holes, we can easily obtain the thermodynamics of $U(1)^2$-charged accelerating black holes. And the strategies are similar to previous ones, so we will not explain too much in this section. The bulk action is
\begin{align}
S_{\text{bulk}}&=\frac{1}{16 \pi G_4}\int_{\mathcal M}\!d^{3+1}x \sqrt{-G}\biggl(R-\frac{1}{2}(\partial\phi)^2-V(\phi)-\frac{1}{4}\sum\limits_{I=1}\limits^{2}Z_I(\phi)\mathcal F_I^2\biggr),
\end{align}
where $\mathcal F_I=dA_I$, $Z_1(\phi)=e^{\phi/\zeta}$, $Z_2(\phi)=e^{-\zeta \phi}$, and the scalar potential is the same as~\eqref{VV}. When $\zeta=1/\sqrt{3}$, $1$, $\sqrt{3}$, these theories can be embedded in gauged STU supergravities.

These theories admit the $U(1)^2$-charged accelerating black hole solutions as follows \cite{Lu:2014sza}:
\begin{equation}
ds^2=\frac{1}{(1+arx)^2}\biggl[f(x)^j\biggl(-\frac{F(r)}{\alpha^2}dt^2+\frac{dr^2}{F(r)}\biggr)+r^2h(r)^j\biggl(\frac{dx^2}{G(x)}+\frac{G(x)d\varphi^2}{K^2}\biggr)\biggr],
\end{equation}
with
\begin{align}
&j=\frac{2\zeta^2}{1+\zeta^2},\qquad h(r)=1-\frac{q}{r},\qquad f(x)=1+aqx,\nonumber\\
&F(r)=(1-a^2r^2)\Big(1-\frac{2m}{r}+\frac{e^2}{r^2}\Big)h(r)^{-j}+\frac{r^2}{\ell^2}h(r)^j,\nonumber\\
&G(x)=(1-x^2)(1+2amx+a^2e^2x^2)f(x)^{-j},\\
&A_1=\frac{\sqrt{2j(1-a^2q^2)\left(e^2-q(2m-q)\right)}}{\alpha}\left(\frac{1}{r_hh(r_h)}-\frac{1}{rh(r)}\right)dt,\nonumber\\
&A_2=\frac{e\sqrt{2(2-j)}}{\alpha}\left(\frac{1}{r_h}-\frac{1}{r}\right)dt,\quad \phi=\sqrt{j(2-j)}\,\text{log}\left(\frac{h(r)}{f(x)}\right).\nonumber
\end{align}
When $A_1=0$, these solutions reduce to accelerating charged dilaton black holes. The limit $q\to0$ takes these solutions to RN-AdS accelerating black hole.

The FG expansion~\eqref{FGT} of these solutions leads to the boundary metric as follows:
\begin{equation}
ds^2_{(0)}=-\frac{\gamma^2_q(\rho)}{\ell^4(1+aq\rho)^{4j}}dt^2+\frac{\alpha^2}{\ell^2X_q(\rho)\gamma_q(\rho)^2}d\rho^2+\frac{\alpha^2X_q(\rho)}{K^2\ell^2(1+aq\rho)^{4j}}d\varphi^2,
\end{equation}
where
\begin{align}
X_q(\rho)&=(1-\rho^2)(1+2am\rho+a^2e^2\rho^2),\\
\gamma_q(\rho)&=\sqrt{(1+aq\rho)^{2j} -a^2\ell^2X_q(\rho)}. 
\end{align}
The stress-energy tensor~\eqref{TmumuT} is [the $x_1$ is the same as in~\eqref{x11}]
\begin{equation}
\langle\mathcal T_{i}^{j}\rangle_\text{T}=\langle\mathcal T_{i}^{j}\rangle_\text{C}+\frac{\ell^4(1+aq\rho)^{4j-3}aj(2-j)\omega_qq^2}{24\pi G_4\alpha^3}\text{diag}(1,1,1),
\end{equation}
and the stress-energy tensor~\eqref{TUN} is
\begin{equation}
\langle\mathcal T_{i}^{j}\rangle_\text{C}=\frac{\ell^4(1+aq\rho)^{2j-3}\omega_e}{48\pi G_4\alpha^3}\text{diag}\left[-\left(2\gamma_q^2-a^2\ell^2X_q\right),(1+aq\rho)^{2j},\left(\gamma_q^2-2a^2\ell^2X_q\right)\right],
\end{equation}
where
\begin{align}
\omega_4&=(1-a^2e^2)j+6a(j-1)m\rho+6a^2e^2(j-2)\rho^2,\nonumber\\
\omega_5&=j(2j-1)\rho+am\left(3(j-2)(2j-3)\rho^2-j(2j+5)\right)\nonumber\\
        &+a^2e^2\rho\left(j-2j^2+2(j-2)(2j-9)\rho^2\right),\\
\omega_6&=2j(1+j)+am\rho\left(j(2j-1)+(j-2)(2j-3)\rho^2\right)+2a^2e^2(j-3)(j-2)\rho^4,\nonumber\\
\omega_e&=6m+12ae^2\rho-3q\omega_4+aq^2\omega_5+a^2q^3\omega_6,\nonumber\\
\omega_q&=\rho+a\left[q-ae^2\rho+ae^2\rho^3(2+aq\rho)+m\left(3\rho^2-1+aq\rho(1+\rho^2)\right)\right].\nonumber
\end{align}
We can find that these two stress-energy tensors yield the same correct mass as the conformal method~\eqref{eq:AMD}, which means that the mass integration~\eqref{massintegrate} related to the $\omega_q$ term is zero. With these results, we can quickly check that the trace of the $\langle\mathcal T_{i}^{j}\rangle_\text{C}$ is zero but the trace of the $\langle\mathcal T_{i}^{j}\rangle_\text{T}$ is not. We can also check that in the limit $q\to0$ these two results reduce to the corresponding result in \cite{Anabalon:2018qfv}. All of these are similar to the $U(1)$-charged solutions, except for the differences of the dictionary~\eqref{TmumuT} and the local Ward identities~\eqref{WDIT},~\eqref{TUN11} are only one more vector field involved. But we are not going to write down the asymptotic expansions of the gauge and scalar field here.

The thermodynamic quantities of these solutions are
\begin{align}\label{THHH}
\Xi&=1+a^2e^2, \qquad\Theta=\frac{e^2-q(2m-q)}{1-a^2q^2}, \qquad\mu_\pm=\frac{1}{4G_4}\biggl[1-\frac{\Xi\pm 2am}{K(1\pm aq)^{j}}\biggr],\nonumber\\
M&=\frac{2m\Big(1+(j-1)a^2q^2\Big)-jq\Xi}{2\alpha KG_4(1-a^2q^2)}-\frac{a\ell^2\Big(\frac{(\Xi+2am)^2}{(1+aq)^{2j}}-\frac{(\Xi-2am)^2}{(1-aq)^{2j}}\Big)}{8\alpha KG_4},\quad \alpha=\sqrt{a\alpha KG_4M},\nonumber\\
T&=\frac{h(r_h)^{-j}}{2\pi\alpha r_h^2}\biggl[(1-a^2r_h^2)\Big(m-\frac{e^2}{r_h}\Big)+\frac{r_h^2\bigl(r_h+qj(1-a^2r_h^2)-q\bigr)}{\ell^2h(r_h)^{1-2j}(1-a^2r_h^2)}\biggr],\nonumber\\
S&=\frac{\pi r_h^2h(r_h)^j}{KG_4\left(1-a^2r_h^2\right)}, \qquad Q_1=\frac{\sqrt{2j\Theta}}{4KG_4}, \qquad \Phi_1=\frac{\sqrt{2j\Theta}(1-a^2qr_h)}{\alpha r_hh(r_h)},\\
Q_2&=\frac{e\sqrt{2(2-j)}}{4KG_4}, \qquad \Phi_2=\frac{e\sqrt{2(2-j)}}{\alpha r_h},\qquad P=\frac{3}{8\pi\ell^2G_4},\nonumber\\
V&=\frac{4\pi\ell^2}{3\alpha K}\biggl[m-\alpha KMG_4-\frac{je^2}{2r_h}+\frac{(1-a^2qr_h)j\Theta}{2r_hh(r_h)}+\frac{r_h^2\bigl(2r_h-q\left(2-j-(1-2a^2r_h^2)j\right)\bigr)}{2\ell^2h(r_h)^{1-2j}(1-a^2r_h^2)^2}\biggr].\nonumber
\end{align}
With these results, we can check that the Smarr relation~\eqref{Smarr} is satisfied. Although we can construct the first law~\eqref{first} from $\delta F(r_h)=0$~\eqref{FRH} with these thermodynamic quantities, there are no simple expressions for the thermodynamic lengths ($\lambda_\pm$) for arbitrary coupling constant. So we only write down $\lambda_\pm$ for the three cases that can be embedded in gauged STU supergravities.

The thermodynamic lengths for $\zeta=1$, $1/\sqrt{3}$, $\sqrt{3}$ are
\begin{itemize}
\item $\zeta=1$

The $\alpha$ in~\eqref{THHH} and its corresponding $\lambda_\pm$ are as follows:
\begin{align}
\alpha&=\sqrt{\frac{a\Big(2m-q\Xi\Big)\Big(1-a^2\ell^2(\Xi+\frac{q^2}{\ell^2}-2a^2mq)\Big)}{2(1-a^2q^2)^2}},\\
\lambda_\pm&=\frac{1}{2\alpha(1\pm aq)}\Big[\frac{ar_h(1\mp ar_h)(2\pm ar_h)\mp1}{a(1-a^2r_h^2)}\mp a\ell^2\Big(1\pm ar_h\Big)\Big(1\pm\frac{ae^2}{r_h}\Big)\nonumber\\
&-q\frac{2\pm a\Big(q-4r_h+aqr_h(ar_h\mp1)\Big)}{1-a^2r_h^2}\Big].\nonumber
\end{align}

If we have a better $\alpha$, then there exists a good limit and its corresponding thermodynamic lengths are
\begin{align}
\alpha&=\frac{\sqrt{\Big(\Xi-2a^2mq\Big)\Big(1-a^2\ell^2(\Xi+\frac{q^2}{\ell^2}-2a^2mq)\Big)}}{1-a^2q^2},\\
\lambda_\pm&=\frac{Y_{0\pm}\pm(Y_1+Y_{1\pm})q+(Y_{2\pm}\mp Y_2)q^2+Y_{3\pm}q^3+Y_{4\pm}q^4+Y_{5\pm}q^5-Y_{6}q^6}{\alpha Y(1-a^2q^2)},\nonumber
\end{align}
where
\begin{align}
&Y=2r_h^2(1-a^2r_h^2)\Big[\ell^2(1-a^2r_h^2)\Big(1+a^2(e^2h(r_h)-qr_h)\Big)-a^2r_h^3qh(r_h)^2\Big],\nonumber\\
&Y_{\pm}=\ell^2r_h\Big(ar_h(ar_h\mp2\Xi)+3\Xi-2\Big)\mp2a\ell^4(1-a^2r_h^2)\Xi^2,\nonumber\\
&Y_{0\pm}=-r_h(1-a^2r_h^2)\Big(e^2\ell^2+r_h(r_h^3-Y_{\pm})\Big),\quad Y_{1}=4a^3\ell^4\Xi r_h(e^2+r_h^2)(1-a^2r_h^2)^2,\nonumber\\
&Y_{1\pm}=2a\ell^2r_h^2(1-a^2r_h^2)\Big(r_h+ar_h^2(3ar_h\mp2)+ae^2(2ar_h\mp)(1+a^2r_h^2)\Big)\nonumber\\
     &\hspace{0.04\textwidth}\pm2r_h^4(1\mp ar_h)\Big(1\pm ar_h(1\mp ar_h)\Big),\quad Y_{2}=2a^5\ell^4(e^2+r_h^2)^2(1-a^2r_h^2)^2,\nonumber\\
&Y_{2\pm}=\mp2a^2\ell^2r_h(1-a^2r_h^2)\Big\{2ar_h^3(2+a^2r_h^2)+e^2\Big[ar_h\Big(1+ar_h(5ar_h\pm1)\Big)\mp1\Big]\Big\}\nonumber\\
       &\hspace{0.04\textwidth}+r_h^3\Big[a^2r_h^2\Big(7\mp2ar_h(3+a^2r_h^2)\Big)-1\Big],\\
&Y_{3\pm}=2a^2r_h^2\Big\{a\ell^2(1-a^2r_h^2)\Big[ar_h^2(2\pm3ar_h)\pm r_h+ae^2\Big(1+ar_h(ar_h\pm4)\Big)\Big]\nonumber\\
       &\hspace{0.04\textwidth}\pm r_h^2\Big[ar_h\Big(3+ar_h(3ar_h\pm1)\Big)\mp3\Big]\Big\},\nonumber\\
&Y_{4\pm}=a^2r_h\Big\{a^2\ell^2(ar_h\mp1)(ar_h\pm1)^3(e^2+r_h^2)\nonumber\\
       &\hspace{0.04\textwidth}+r_h^2\Big[2\mp ar_h\Big(2+ar_h(1\pm ar_h)(ar_h\pm5)\Big)\Big]\Big\},\nonumber\\
&Y_{5\pm}=2a^4r_h^4\Big(2+ar_h(ar_h\pm1)\Big),\qquad Y_{6}=a^4r_h^3(1+a^2r_h^2).\nonumber
\end{align}
We can check that the limit $q\to0$ takes this $\alpha$ to $\sqrt{\Xi(1-a^2\ell^2\Xi)}$, which is the same as in \cite{Anabalon:2018qfv} for an accelerating RN-AdS black hole. The limits $e\to\sqrt{q(2m-q)}$ and $2m\to2c+q$ take this result to~\eqref{THQ1}.

\item $\zeta=1/\sqrt{3}$
\begin{align}
\alpha&=\sqrt{\frac{a\Big[4m\Big(1-a^2\ell^2(\Xi+\frac{q^2}{2\ell^2}-a^2mq)\Big)-\Xi(1-a^2\ell^2\Xi)q\Big]}{4(1-a^2q^2)}},\\
\lambda_\pm&=\frac{1}{2\alpha\sqrt{1\pm ab}}\left(r_h-a^2\ell^2(r_h+\frac{e^2}{r_h})\pm\frac{1-a^2\ell^2\Xi}{a}+\frac{ar_h\mp2}{a(1-a^2r_h^2)}-\frac{q(1\mp2ar_h)}{1-a^2r_h^2}\right).\nonumber
\end{align}

\item $\zeta=\sqrt{3}$
\begin{align}
\alpha&=\sqrt{\frac{a\Big(2m(2+a^2q^2)-3q\Xi\Big)}{4(1-a^2q^2)}-\frac{1}{8}a^2\ell^2\Big(\frac{(\Xi+2am)^2}{(1+aq)^3}-\frac{(\Xi-2am)^2}{(1-aq)^3}\Big)},\nonumber\\
\lambda_\pm&=\frac{1}{2\alpha\sqrt{(1\pm ab)^3}}\Big[r_h-a^2\ell^2(r_h+\frac{e^2}{r_h})\pm\frac{1-a^2\ell^2\Xi}{a}+\frac{ar_h\mp2}{a(1-a^2r_h^2)}\\
           &+q\frac{ar_h\Big(aq(3\pm2aq)\pm6\Big)-3(1\pm aq)-qa^3r_h^2(aq\pm3)}{1-a^2r_h^2}\Big].\nonumber
\end{align}
\end{itemize}
When $\zeta=1$, the conditions for supersymmetry and extremality of $U(1)^2$-magnetically charged accelerating black hole have been studied in \cite{Ferrero:2021ovq}, in which the on-shell action and the mass were guessed. It would be very straightforward to apply the methods we use to their model. However, since the equations of motion of this system are unchanging under the electric-magnetic duality transformation and $\phi\to-\phi$, the thermodynamics of a magnetically charged one is similar to the electrically charged one. There is no doubt that one can analyze the $U(1)^4$-electrically or magnetically charged accelerating black holes \cite{Lu:2014sza} (see also \cite{Ferrero:2021ovq}) along this line. 

\section{Summary and Discussion}

To summarize, we have derived the thermodynamic laws for slowly accelerating charged dilaton black holes in four-dimensional Einstein-Maxwell-dilaton theories for arbitrary dilaton coupling constant. These systems can be embedded in gauged supergravities when $\zeta=0$, $1/\sqrt{3}$, $1$, and $\sqrt{3}$. The thermodynamics of slowly accelerating black holes is generalized to include scalar hair, which does not introduce a scalar charge. Like accelerating black holes without scalar hair \cite{Podolsky:2003gm, Podolsky:2009ag}, the hairy accelerating black holes emit gravitational and electromagnetic radiations at infinity. The standard thermodynamic analysis is still valid, and the thermodynamic quantities we obtained are related to the comoving observer, who cannot receive any radiation. We compute the holographic stress-energy tensors and other one-point functions by properly considering mixed boundary conditions for the scalar field. The stress-energy tensor describes a thermal perfect fluid plus a nonhydrodynamic correction \cite{Anabalon:2018ydc, Anabalon:2018qfv, BernardideFreitas:2014eoi}. We find that the holographic mass associated with appropriate boundary conditions for the scalar field agrees with the AMD mass. These results can be straightforwardly generalized to $U(1)^2$-charged accelerating black holes.

Our results open a way to analyze the phase structure of slowly accelerating black holes with gauge and scalar fields in asymptotically AdS spacetimes. Even for the nonaccelerating black hole in the $\zeta=1$ case, rich phase structures (e.g., zeroth order phase transition) were obtained \cite{Caldarelli:2016nni}. For general $\zeta$, thermodynamic instabilities for nonaccelerating charged dilaton black holes were studied in \cite{Sheykhi:2009pf}, for example. Thus, a general accelerating hairy black hole is expected to have rather richer phases.

A key point of our work is rescaling the time coordinate with the parameter $\alpha$. However, the $\alpha$ we obtained for general $\zeta$ does not have a good limit when $a\to 0$. We find a better $\alpha$ for the $\zeta=1$ solution, and the better $\alpha$ for arbitrary $\zeta$ is still an unsolved problem. A similar situation is that in the limit $\ell\to\infty$, the thermodynamics of accelerating black holes in AdS cannot be reduced to the result in asymptotically flat spacetime. The thermodynamics of pair creation of dilaton black holes in asymptotically flat spacetime \cite{Dowker:1993bt} needs further study.

There is another subtlety on the consistency of the AMD mass and the holographic mass. According to \cite{Anabalon:2014fla}, they are not the same when the conformal symmetry is broken on the AdS boundary.
In our cases, although the conformal symmetry is broken for the triple-trace deformation boundary condition ($\langle\mathcal T^i{}_i\rangle_\text{T}\neq0$), the AMD mass is the same as the holographic mass. However, we also find a conformal boundary condition that preserves the conformal symmetry for these solutions we analyze. One may compare these masses with the boost mass \cite{Dutta:2005iy}.

For slowly accelerating black holes, the constraints $\gamma^2_b(\rho )\textgreater 0$ and $\gamma^2_q(\rho )\textgreater 0$ in the parameter space ensures the absence of accelerating horizons. If these conditions are not satisfied, then the black funnels and droplets \cite{Hubeny:2009kz, Caldarelli:2011wa} with scalar hair will appear, and we expect that properly taking into account the parameter restrictions can also take these systems to thermal equilibrium. It would be interesting to construct black funnels and droplets with scalar hair.

\acknowledgments
We would like to thank Kostas Skenderis for general comments. This work is supported in part by the NSF of China under Grant No. 11905298.

\appendix

\section{The special cases of gauged supergravity}
The consistent $S^7$ reduction of $11$-dimensional supergravity gives the Lagrangian as follows \cite{Cvetic:1999xp} (see also \cite{Lu:2014sza}):
\begin{equation}\label{Lagrangian}
\mathcal L=R-\frac{1}{2}\sum_{i=1}^3(\partial\phi_i)^2-\frac{1}{4}\sum_{I=1}^4e^{\vec{a}_I\cdot\vec{\phi}}\mathcal F_I^2+\frac{2}{\ell^2}\sum_{i=1}^3\text{cosh}\,\phi_i\,,
\end{equation}
where
\begin{equation}
\vec{a}_1=(1,1,1),\qquad \vec{a}_2=(1,-1,-1),\qquad \vec{a}_3=(-1,1,-1),\qquad \vec{a}_4=(-1,-1,1)\,.
\end{equation}
The $U(1)$-charged Lagrangians are obtained by consistent truncation of~\eqref{Lagrangian}:
\begin{itemize}
\item The $\zeta=0$ case (RN-AdS): $\mathcal F_1=\mathcal F$, $\mathcal F_2=\mathcal F_3=\mathcal F_4=\phi_i=0$. The Lagrangian is
\begin{equation}
\mathcal L=R-\frac{1}{4}\mathcal F^2+\frac{6}{\ell^2}\,.
\end{equation}

\item The $\zeta=1/\sqrt{3}$ case: $\mathcal F_2=\mathcal F$, $\phi_i=\phi/\sqrt{3}$, $\mathcal F_1=\mathcal F_3=\mathcal F_4=0$. The Lagrangian is
\begin{equation}
\mathcal L=R-\frac{1}{2}(\partial\phi)^2-\frac{1}{4}e^{-\frac{\phi}{\sqrt{3}}}\mathcal F^2+\frac{6}{\ell^2}\text{cosh}\,\frac{\phi}{\sqrt{3}}\,.
\end{equation}

\item The $\zeta=1$ case: $\mathcal F_3=\mathcal F$, $\phi_1=\phi$, $\mathcal F_1=\mathcal F_2=\mathcal F_4=\phi_2=\phi_3=0$. The Lagrangian is
\begin{equation}
\mathcal L=R-\frac{1}{2}(\partial\phi)^2-\frac{1}{4}e^{-\phi}\mathcal F^2+\frac{2}{\ell^2}(\text{cosh}\,\phi+2)\,.
\end{equation}

\item The $\zeta=\sqrt{3}$ case: $\mathcal F_1=\mathcal F$, $\phi_i=-\phi/\sqrt{3}$, $\mathcal F_2=\mathcal F_3=\mathcal F_4=0$. The Lagrangian is
\begin{equation}
\mathcal L=R-\frac{1}{2}(\partial\phi)^2-\frac{1}{4}e^{-\sqrt{3}\phi}\mathcal F^2+\frac{6}{\ell^2}\text{cosh}\,\frac{\phi}{\sqrt{3}}\,.
\end{equation}
\end{itemize}
The $U(1)^2$-charged Lagrangians are as follows:
\begin{itemize}
\item The $\zeta=1/\sqrt{3}$ case: $\phi_i=\phi/\sqrt{3}$, $\mathcal F_3=\mathcal F_4=0$. The Lagrangian is
\begin{equation}
\mathcal L=R-\frac{1}{2}(\partial\phi)^2-\frac{1}{4}e^{\sqrt{3}\phi}\mathcal F_1^2-\frac{1}{4}e^{-\frac{\phi}{\sqrt{3}}}\mathcal F_2^2+\frac{6}{\ell^2}\text{cosh}\,\frac{\phi}{\sqrt{3}}\,.
\end{equation}

\item The $\zeta=1$ case: $\phi_1=\phi$, $\mathcal F_2=\mathcal F_3=\phi_2=\phi_3=0$. The Lagrangian is
\begin{equation}
\mathcal L=R-\frac{1}{2}(\partial\phi)^2-\frac{1}{4}e^{\phi}\mathcal F_1^2-\frac{1}{4}e^{-\phi}\mathcal F_4^2+\frac{2}{\ell^2}(\text{cosh}\,\phi+2)\,.
\end{equation}

\item The $\zeta=\sqrt{3}$ case: $\phi_i=-\phi/\sqrt{3}$, $\mathcal F_2=\mathcal F_3=0$. The Lagrangian is
\begin{equation}
\mathcal L=R-\frac{1}{2}(\partial\phi)^2-\frac{1}{4}e^{-\sqrt{3}\phi}\mathcal F_1^2-\frac{1}{4}e^{\frac{\phi}{\sqrt{3}}}\mathcal F_4^2+\frac{6}{\ell^2}\text{cosh}\,\frac{\phi}{\sqrt{3}}\,.
\end{equation}
\end{itemize}


\begin{thebibliography}{99}



\bibitem{Weyl:1917ur}
H. Weyl,
\emph{On the theory of gravitation},
\doi{10.1002/andp.19173591804}{Ann. Phys. 54 (1917) 117.}


\bibitem{EhlersKundt}
J. Ehlers,  W. Kundt, {\it Exact solutions of the gravitational
field equations}, in {\it Gravitation: an introduction to current
research}, edited by  L. Witten (Wiley, New York, London, 1962).



\bibitem{Kinnersley:1970zw}
W.~Kinnersley and M.~Walker,
\emph{Uniformly accelerating charged mass in general relativity},
\doi{10.1103/PhysRevD.2.1359}{Phys. Rev. D \textbf{2} (1970) 1359}



\bibitem{Plebanski:1976gy}
J.~F.~Plebanski and M.~Demianski,
\emph{Rotating, charged, and uniformly accelerating mass in general relativity},
\doi{10.1016/0003-4916(76)90240-2}{Annals Phys. \textbf{98} (1976) 98}



\bibitem{Dias:2002mi}
O.~J.~C.~Dias and J.~P.~S.~Lemos,
\emph{Pair of accelerated black holes in anti-de Sitter background: AdS C metric},
\doi{10.1103/PhysRevD.67.064001}{Phys. Rev. D \textbf{67} (2003) 064001}
[\arXiv{hep-th/0210065}{hep-th}].



\bibitem{Griffiths:2006tk}
J.~B.~Griffiths, P.~Krtous and J.~Podolsky,
\emph{Interpreting the C-metric},
\doi{10.1088/0264-9381/23/23/008}{Class. Quant. Grav. \textbf{23} (2006) 6745}
[\arXiv{gr-qc/0609056}{gr-qc}].



\bibitem{Emparan:1999wa}
R.~Emparan, G.~T.~Horowitz and R.~C.~Myers,
\emph{Exact description of black holes on branes},
\doi{10.1088/1126-6708/2000/01/007}{JHEP \textbf{01} (2000) 007}
[\arXiv{hep-th/9911043}{hep-th}].



\bibitem{Emparan:1999fd}
R.~Emparan, G.~T.~Horowitz and R.~C.~Myers,
\emph{Exact description of black holes on branes. 2. Comparison with BTZ black holes and black strings},
\doi{10.1088/1126-6708/2000/01/021}{JHEP \textbf{01} (2000) 021}
[\arXiv{hep-th/9912135}{hep-th}].



\bibitem{Emparan:2009dj}
R.~Emparan and G.~Milanesi,
\emph{Exact gravitational dual of a plasma ball},
\doi{10.1088/1126-6708/2009/08/012}{JHEP \textbf{08} (2009) 012}
[\arXiv{0905.4590}{hep-th}].



\bibitem{Hubeny:2009kz}
V.~E.~Hubeny, D.~Marolf and M.~Rangamani,
\emph{Black funnels and droplets from the AdS C-metrics},
\doi{10.1088/0264-9381/27/2/025001}{Class. Quant. Grav. \textbf{27} (2010) 025001}
[\arXiv{0909.0005}{hep-th}].



\bibitem{Caldarelli:2011wa}
M.~M.~Caldarelli, O.~J.~C.~Dias, R.~Monteiro and J.~E.~Santos,
\emph{Black funnels and droplets in thermal equilibrium},
\doi{10.1007/JHEP05(2011)116}{JHEP \textbf{05} (2011) 116}
[\arXiv{1102.4337}{hep-th}].



\bibitem{Emparan:2020znc}
R.~Emparan, A.~M.~Frassino and B.~Way,
\emph{Quantum BTZ black hole},
\doi{10.1007/JHEP11(2020)137}{JHEP \textbf{11} (2020) 137}
[\arXiv{2007.15999}{hep-th}].



\bibitem{Ferrero:2020twa}
P.~Ferrero, J.~P.~Gauntlett, J.~M.~P.~Ipi\~na, D.~Martelli and J.~Sparks,
\emph{Accelerating black holes and spinning spindles},
\doi{10.1103/PhysRevD.104.046007}{Phys. Rev. D \textbf{104} (2021) 046007}
[\arXiv{2012.08530}{hep-th}].



\bibitem{Bekenstein:1973ur}
J.~D.~Bekenstein,
\emph{Black holes and entropy},
\doi{10.1103/PhysRevD.7.2333}{Phys.\ Rev.\ D {\bf7} (1973) 2333.}



\bibitem{Bekenstein:1974ax}
J.~D.~Bekenstein,
\emph{Generalized second law of thermodynamics in black hole physics},
\doi{10.1103/PhysRevD.9.3292}{Phys.\ Rev.\ D {\bf9} (1974) 3292.}



\bibitem{Hawking:1974sw}
S.~W.~Hawking,
\emph{Particle creation by black holes},
\doi{10.1007/BF02345020}{Commun.\ Math.\ Phys.\  {\bf 43}, 199 (1975)}
\doi{10.1007/BF02345020}{Erratum: [Commun.\ Math.\ Phys.\ {\bf 46} (1976) 206].}



\bibitem{Kubiznak:2016qmn}
D.~Kubiz\v{n}\'ak, R.~B.~Mann and M.~Teo,
\emph{Black hole chemistry: thermodynamics with Lambda},
\doi{10.1088/1361-6382/aa5c69}{Class. Quant. Grav. \textbf{34} (2017) 063001}
[\arXiv{1608.06147}{hep-th}].



\bibitem{Appels:2016uha}
M.~Appels, R.~Gregory and D.~Kubiz\v{n}\'ak,
\emph{Thermodynamics of accelerating black holes},
\doi{10.1103/PhysRevLett.117.131303}{Phys.\ Rev.\ Lett. \textbf{117} (2016) 131303}
[\arXiv{1604.08812}{hep-th}].



\bibitem{Astorino:2016xiy}
M.~Astorino,
\emph{CFT duals for accelerating black holes},
\doi{10.1016/j.physletb.2016.07.019}{Phys. Lett. B \textbf{760} (2016) 393}
[\arXiv{1605.06131}{hep-th}].



\bibitem{Astorino:2016ybm}
M.~Astorino,
\emph{Thermodynamics of regular accelerating black holes},
\doi{10.1103/PhysRevD.95.064007}{Phys. Rev. D \textbf{95} (2017) 064007}
[\arXiv{1612.04387}{gr-qc}].



\bibitem{Appels:2017xoe}
M.~Appels, R.~Gregory and D.~Kubiz\v{n}\'ak,
\emph{Black hole thermodynamics with conical defects},
\doi{10.1007/JHEP05(2017)116}{JHEP \textbf{05} (2017) 116}
[\arXiv{1702.00490}{hep-th}].



\bibitem{Anabalon:2018ydc}
A.~Anabal\'on, M.~Appels, R.~Gregory, D.~Kubiz\v{n}\'ak, R.~B.~Mann and A.~Ovg\"un,
\emph{Holographic thermodynamics of accelerating black holes},
\doi{10.1103/PhysRevD.98.104038}{Phys. Rev. D \textbf{98} (2018) 104038}
[\arXiv{1805.02687}{hep-th}].



\bibitem{Anabalon:2018qfv}
A.~Anabal\'on, F.~Gray, R.~Gregory, D.~Kubiz\v{n}\'ak and R.~B.~Mann,
\emph{Thermodynamics of charged, rotating, and accelerating black holes},
\doi{10.1007/JHEP04(2019)096}{JHEP \textbf{04} (2019) 096}
[\arXiv{1811.04936}{hep-th}].



\bibitem{Gibbons:2004ai}
G.~W.~Gibbons, M.~J.~Perry and C.~N.~Pope,
\emph{The First law of thermodynamics for Kerr-anti-de Sitter black holes},
\doi{10.1088/0264-9381/22/9/002}{Class. Quant. Grav. \textbf{22} (2005) 1503}
[\arXiv{hep-th/0408217}{hep-th}].



\bibitem{Cassani:2021dwa}
D.~Cassani, J.~P.~Gauntlett, D.~Martelli and J.~Sparks,
\emph{Thermodynamics of accelerating and supersymmetric AdS4 black holes},
\doi{10.1103/PhysRevD.104.086005}{Phys. Rev. D \textbf{104} (2021) 086005}
[\arXiv{2106.05571}{hep-th}].



\bibitem{Visser:2021eqk}
M.~R.~Visser,
\emph{Holographic thermodynamics requires a chemical potential for color},
\doi{10.1103/PhysRevD.105.106014}{Phys. Rev. D \textbf{105} (2022) 106014}
[\arXiv{2101.04145}{hep-th}].



\bibitem{Cong:2021fnf}
W.~Cong, D.~Kubiz\v{n}\'ak and R.~B.~Mann,
\emph{Thermodynamics of AdS black holes: critical behavior of the Central Charge},
\doi{10.1103/PhysRevLett.127.091301}{Phys. Rev. Lett. \textbf{127} (2021) 091301}
[\arXiv{2105.02223}{hep-th}].



\bibitem{Lu:2014ida}
H.~L\"{u} and J.~F.~Vazquez-Poritz,
\emph{Dynamic C-metrics in gauged supergravities},
\doi{10.1103/PhysRevD.91.064004}{Phys.\ Rev.\ D {\bf 91} (2015) 064004}
[\arXiv{1408.3124}{[hep-th]}].
  
  

\bibitem{Ren:2019lgw}
J.~Ren,
\emph{Analytic solutions of neutral hyperbolic black holes with scalar hair},
\doi{10.1103/PhysRevD.106.086023}{Phys. Rev. D \textbf{106} (2022) 086023}
[\arXiv{1910.06344}{hep-th}].



\bibitem{Ren:2021rhx}
J.~Ren and W.~Zheng,
\emph{Analytic AC conductivities from holography},
\doi{10.1103/PhysRevD.105.066013}{Phys. Rev. D \textbf{105} (2022) 066013}
[\arXiv{2109.07481}{hep-th}].



\bibitem{Dowker:1993bt}
F.~Dowker, J.~P.~Gauntlett, D.~A.~Kastor and J.~H.~Traschen,
\emph{Pair creation of dilaton black holes},
\doi{10.1103/PhysRevD.49.2909}{Phys.\ Rev.\ D {\bf 49} (1994) 2909}
[\arXivid{hep-th}{9309075}].


 
\bibitem{Gregory:1995hd}
R.~Gregory and M.~Hindmarsh,
\emph{Smooth metrics for snapping strings},
\doi{10.1103/PhysRevD.52.5598}{Phys. Rev. D \textbf{52} (1995) 5598}
[\arXiv{gr-qc/9506054}{gr-qc}].



\bibitem{Lu:2014sza}
H.~L\"u and J.~F.~V\'azquez-Poritz,
\emph{C-metrics in gauged STU supergravity and beyond},
\doi{10.1007/JHEP12(2014)057}{JHEP \textbf{12} (2014) 057}
[\arXiv{1408.6531}{hep-th}].



\bibitem{Ferrero:2021ovq}
P.~Ferrero, M.~Inglese, D.~Martelli and J.~Sparks,
\emph{Multicharge accelerating black holes and spinning spindles},
\doi{10.1103/PhysRevD.105.126001}{Phys. Rev. D \textbf{105} (2022) 126001}
[\arXiv{2109.14625}{hep-th}].



\bibitem{Benini:2015eyy}
F.~Benini, K.~Hristov and A.~Zaffaroni,
\emph{Black hole microstates in AdS$_{4}$ from supersymmetric localization},
\doi{10.1007/JHEP05(2016)054}{JHEP \textbf{05} (2016) 054}
[\arXiv{1511.04085}{hep-th}].



\bibitem{Cabo-Bizet:2018ehj}
A.~Cabo-Bizet, D.~Cassani, D.~Martelli and S.~Murthy,
\emph{Microscopic origin of the Bekenstein-Hawking entropy of supersymmetric AdS$_{5}$ black holes},
\doi{10.1007/JHEP10(2019)062}{JHEP \textbf{10} (2019) 062}
[\arXiv{1810.11442}{hep-th}].



\bibitem{Podolsky:2002nk}
J.~Podolsky,
\emph{Accelerating black holes in anti-de Sitter universe},
\doi{10.1023/A:1013961411430}{Czech. J. Phys. \textbf{52} (2002) 1}
[\arXiv{gr-qc/0202033}{gr-qc}].



\bibitem{Gibbons:1996af}
G.~W.~Gibbons, R.~Kallosh and B.~Kol,
\emph{Moduli, scalar charges, and the first law of black hole thermodynamics},
\doi{10.1103/PhysRevLett.77.4992}{Phys. Rev. Lett. \textbf{77} (1996) 4992}
[\arXiv{hep-th/9607108}{hep-th}].



\bibitem{Lu:2013ura}
H.~L\"u, Y.~Pang and C.~N.~Pope,
\emph{AdS dyonic black hole and its thermodynamics},
\doi{10.1007/JHEP11(2013)033}{JHEP \textbf{11} (2013) 033}
[\arXiv{1307.6243}{hep-th}].



\bibitem{Lu:2014maa}
H.~L\"{u}, C.~N.~Pope and Q.~Wen,
\emph{Thermodynamics of AdS black holes in Einstein-Scalar Gravity},
\doi{10.1007/JHEP03(2015)165}{JHEP \textbf{03} (2015) 165}
[\arXiv{1408.1514}{hep-th}].



\bibitem{Caldarelli:2016nni}
M.~M.~Caldarelli, A.~Christodoulou, I.~Papadimitriou and K.~Skenderis,
\emph{Phases of planar AdS black holes with axionic charge},
\doi{10.1007/JHEP04(2017)001}{JHEP \textbf{04} (2017) 001}
[\arXiv{1612.07214}{hep-th}].



\bibitem{Astefanesei:2018vga}
D.~Astefanesei, R.~Ballesteros, D.~Choque and R.~Rojas,
\emph{Scalar charges and the first law of black hole thermodynamics},
\doi{10.1016/j.physletb.2018.05.005}{Phys. Lett. B \textbf{782} (2018) 47}
[\arXiv{1803.11317}{hep-th}].



\bibitem{Smarr:1972kt}
L.~Smarr,
\emph{Mass formula for Kerr black holes},
\doi{10.1103/PhysRevLett.30.71}{Phys. Rev. Lett. \textbf{30} (1973) 71}
\doi{10.1103/PhysRevLett.30.71}{[erratum: Phys. Rev. Lett. \textbf{30} (1973), 521]}


\bibitem{Kastor:2009wy}
D.~Kastor, S.~Ray and J.~Traschen,
\emph{Enthalpy and the mechanics of AdS black holes},
\doi{10.1088/0264-9381/26/19/195011}{Class. Quant. Grav. \textbf{26} (2009) 195011}
[\arXiv{0904.2765}{hep-th}].



\bibitem{Gao:2004tu}
C.~J.~Gao and S.~N.~Zhang,
\emph{Dilaton black holes in de Sitter or Anti-de Sitter universe},
\doi{10.1103/PhysRevD.70.124019}{Phys. Rev. D \textbf{70} (2004) 124019}
[\arXiv{hep-th/0411104}{hep-th}].



\bibitem{FG}
C.~Fefferman and C.~Robin Graham,
{\it Conformal invariants} in
{\it Elie Cartan et les Math\'ematiques d'aujourd 'hui}
(Ast\'erisque, 1985) Numero Hors Serie, 95-116, see also:
[\arXiv{0710.0919}{math.DG}].



\bibitem{Witten:2001ua}
E.~Witten,
\emph{Multitrace operators, boundary conditions, and AdS/CFT correspondence},
[\arXiv{hep-th/0112258}{hep-th}].



\bibitem{Mueck:2002gm}
W.~Mueck,
\emph{An improved correspondence formula for AdS/CFT with multitrace operators},
\doi{10.1016/S0370-2693(02)01487-9}{Phys. Lett. B \textbf{531} (2002) 301}
[\arXiv{hep-th/0201100}{hep-th}].



\bibitem{Papadimitriou:2007sj}
I.~Papadimitriou,
\emph{Multi-trace deformations in AdS/CFT: exploring the vacuum structure of the deformed CFT},
\doi{10.1088/1126-6708/2007/05/075}{JHEP \textbf{05} (2007) 075}
[\arXiv{hep-th/0703152}{hep-th}].



\bibitem{Ashtekar:1984zz}
A.~Ashtekar and A.~Magnon,
\emph{Asymptotically anti-de Sitter space-times},
\doi{10.1088/0264-9381/1/4/002}{Class. Quant. Grav. \textbf{1} (1984) L39}



\bibitem{Ashtekar:1999jx}
A.~Ashtekar and S.~Das,
\emph{Asymptotically anti-de Sitter space-times: Conserved quantities},
\doi{10.1088/0264-9381/17/2/101}{Class. Quant. Grav. \textbf{17} (2000) L17}
[\arXiv{hep-th/9911230}{hep-th}].



\bibitem{Herdeiro:2009vd}
C.~Herdeiro, B.~Kleihaus, J.~Kunz and E.~Radu,
\emph{On the Bekenstein-Hawking area law for black objects with conical singularities},
\doi{10.1103/PhysRevD.81.064013}{Phys. Rev. D \textbf{81} (2010) 064013}
[\arXiv{0912.3386}{gr-qc}].



\bibitem{Hawking:1995ap}
S.~W.~Hawking and S.~F.~Ross,
\emph{Duality between electric and magnetic black holes},
\doi{10.1103/PhysRevD.52.5865}{Phys. Rev. D \textbf{52} (1995) 5865}
[\arXiv{hep-th/9504019}{hep-th}].



\bibitem{Podolsky:2003gm}
J.~Podolsky, M.~Ortaggio and P.~Krtous,
\emph{Radiation from accelerated black holes in an anti-de Sitter universe},
\doi{10.1103/PhysRevD.68.124004}{Phys. Rev. D \textbf{68} (2003) 124004}
[\arXiv{gr-qc/0307108}{gr-qc}].



\bibitem{Podolsky:2009ag}
J.~Podolsky and H.~Kadlecova,
\emph{Radiation generated by accelerating and rotating charged black holes in (anti-)de Sitter space},
\doi{10.1088/0264-9381/26/10/105007}{Class. Quant. Grav. \textbf{26} (2009) 105007}
[\arXiv{0903.3577}{gr-qc}].



\bibitem{BernardideFreitas:2014eoi}
G.~Bernardi de Freitas and H.~S.~Reall,
\emph{Algebraically special solutions in AdS/CFT},
\doi{10.1007/JHEP06(2014)148}{JHEP \textbf{06} (2014) 148}
[\arXiv{1403.3537}{hep-th}].


\bibitem{Sheykhi:2009pf}
A.~Sheykhi, M.~H.~Dehghani and S.~H.~Hendi,
\emph{Thermodynamic instability of charged dilaton black holes in AdS spaces},
\doi{10.1103/PhysRevD.81.084040}{Phys. Rev. D \textbf{81} (2010) 084040}
[\arXiv{0912.4199}{hep-th}].



\bibitem{Anabalon:2014fla}
A.~Anabalon, D.~Astefanesei and C.~Martinez,
\emph{Mass of asymptotically anti\textendash{}de Sitter hairy spacetimes},
\doi{10.1103/PhysRevD.91.041501}{Phys. Rev. D \textbf{91} (2015) 041501}
[\arXiv{1407.3296}{hep-th}].



\bibitem{Dutta:2005iy}
K.~Dutta, S.~Ray and J.~Traschen,
\emph{Boost mass and the mechanics of accelerated black holes},
\doi{10.1088/0264-9381/23/2/005}{Class. Quant. Grav. \textbf{23} (2006) 335}
[\arXiv{hep-th/0508041}{hep-th}].



\bibitem{Cvetic:1999xp}
M.~Cvetic, M.~J.~Duff, P.~Hoxha, J.~T.~Liu, H.~Lu, J.~X.~Lu, R.~Martinez-Acosta, C.~N.~Pope, H.~Sati and T.~A.~Tran,
\emph{Embedding AdS black holes in ten-dimensions and eleven-dimensions},
\doi{10.1016/S0550-3213(99)00419-8}{Nucl. Phys. B \textbf{558} (1999), 96}
[\arXiv{hep-th/9903214}{hep-th}].



\end{thebibliography}
\end{document}